\documentclass[authoryear, preprint, 12pt]{elsarticle}
\usepackage{amsmath}
\usepackage{amssymb}
\usepackage{booktabs}
\usepackage{float}
\usepackage{url}
\usepackage[protrusion=true,expansion=false]{microtype}
\usepackage[
  colorlinks=true,
  linkcolor={black!70!blue},
  citecolor={black!60!green},
  urlcolor={black!70!blue},
  pdftitle={Development and Evaluation of a Sequence-to-Sequence ConvLSTM Approach for Leaf Area Index Forecasting over the South-Central United States},
  pdfauthor={Zhixing Ruan, Lixin Lu},
  pdfsubject={Leaf Area Index forecasting with ConvLSTM sequence-to-sequence models},
  pdfkeywords={Leaf Area Index, subseasonal forecasting, ConvLSTM, deep learning, MODIS, meteorological forcing}
]{hyperref}

\begin{document}

\begin{frontmatter}

\title{\texorpdfstring{\large }{}A Sequence-to-Sequence ConvLSTM Approach for Leaf Area Index Forecasting over the South-Central United States} 

\author[1]{Zhixing Ruan}
\ead{zhixing.ruan@colostate.edu}
\author[1]{Lixin Lu}
\ead{lixin.lu@colostate.edu}
\affiliation[1]{
    organization={Cooperative Institute for Research in the Atmosphere, Colorado State University \\ Fort Collins, Colorado},
    country={USA}}

\begin{abstract}
Leaf Area Index (LAI) is a fundamental biophysical variable governing land-atmosphere interactions; however, LAI forecasting at high spatial resolution remains an unsolved challenge. While recent machine learning approaches have demonstrated LAI estimation at point or regional scales, none provides a gridded, meteorology-driven prognostic forecast suitable for subseasonal land surface and climate modeling applications. Here we present a sequence-to-sequence Convolutional LSTM (ConvLSTM) framework that generates daily 1-km LAI forecasts up to 30 days ahead, driven by historical LAI sequences and daily meteorological forcing including temperature and precipitation. Trained and evaluated over the South-Central United States --- a region of strong climate gradients and diverse vegetation --- the model achieves a domain-averaged RMSE of 0.36 at a 30-day lead time, more than a third lower than the persistence baseline. Forecast skill remains robust across seasons, geographic distributions, and plant functional types, including forests, grasslands, shrublands, and croplands. To our knowledge, this is the first demonstration of skillful LAI forecasting at a 30-day horizon at 1-km resolution.
\end{abstract}



\end{frontmatter}

\clearpage
\noindent\textbf{Highlights}
\begin{itemize}
\item A sequence-to-sequence ConvLSTM forecasts gridded daily LAI up to 30 days ahead with meteorological predictors
\item Forecasts achieve RMSE of 0.36 and R$^2$ of 0.82 at a 30-day lead time, $\sim$35\% below the persistence baseline
\item Forecast skill is assessed by season, plant functional type, and bias structure, beyond aggregate metrics
\item Seasonal LAI bias aligns with precipitation anomalies, linking forecast error to meteorological variability
\item Incorporating land cover improves forecasts for vegetation with distinctive phenology (shrublands, croplands)
\end{itemize}

\section{Introduction}
\label{sec1}

Vegetation plays a crucial role in regulating both regional and global climate systems. Rather than serving as passive landscape cover, the biosphere dynamically modulates surface energy partitioning, the hydrological cycle, and carbon fluxes through processes such as transpiration, albedo modification, and stomatal regulation \citep{bonan2008}. These vegetation dynamics exert a complex, often non-linear influence on atmospheric processes, including temperature, precipitation, and boundary layer development \citep{pielke1998}. Coupled atmosphere-vegetation modeling has demonstrated that vegetation phenology responds strongly to atmospheric forcing on weekly to annual timescales, and that representing vegetation state in climate models significantly improves simulated regional climate \citep{lu2001}. 

Leaf area index (LAI), defined as the one-sided green leaf area per unit ground area, is a key biophysical variable integrating vegetation structure and physiological state \citep{chen1992}. Assimilating satellite-derived LAI into regional climate models has been shown to produce more realistic simulations of surface temperature and precipitation compared to prescribed vegetation parameters \citep{lu2002}. However, operational forecasting requires prognostic LAI rather than observed retrievals, and accurately predicting future vegetation states remains challenging. While Earth System Models attempt to simulate future vegetation states, their LAI representations often struggle to capture mean state, seasonal cycles, and interannual variability \citep{mahowald2016}. In addition, running a fully interactive phenology module inside a climate model is computationally expensive. These limitations motivate the development of an efficient machine learning (ML) approach to prognostic LAI modeling.

Several studies have applied traditional ML models to simulate LAI from meteorological or remote sensing inputs, including support vector regression \citep{zhang2021}, random forest \citep{dong2025,du2022}, and gradient boosting methods \citep{wang2023}. However, these traditional ML models lack temporal forecasting capability, as their point-to-point static mapping fails to capture the cumulative, time-lagged memory of the land surface. Recent deep learning approaches, including attention-enhanced Long Short-Term Memory (LSTM) \citep{xiong2024} and CNN-LSTM hybrids \citep{peng2026}, have shown the capacity to capture temporal dependencies in vegetation modeling. Yet, because these methods require concurrent meteorological observations as inputs, they function primarily as diagnostic estimations rather than prognostic forecasting models. Convolutional LSTM (ConvLSTM) extends the standard LSTM with convolutional operations to jointly capture spatial and temporal dependencies in grid-structure data \citep{shi2015}. \citet{kartal2024} applied ConvLSTM to forecast the Vegetation Health Index (VHI) from MODIS time series, achieving RMSE as low as 0.025 for 8-day ahead predictions using only historical VHI images. However, their model relies solely on the autoregressive history of the target variable and incorporates no meteorological forcing, limiting its ability to predict vegetation responses to weather and climate variations.

Existing satellite-derived LAI products provide accurate retrospective estimates but lack forecasting capability, while current ML approaches either treat vegetation variables as a static mapping target or rely on autoregressive history without meteorological forcing. Yet atmospheric conditions are known to be primary drivers of LAI variability: \citet{Lu1999} demonstrated, based on coupled atmospheric-ecological modeling experiments over the central United States, that precipitation is the dominant driver of LAI variability, with maximum and minimum temperature playing a secondary but notable role. Incorporating such forcing variables is critical in the South-Central United States, a transition zone where rapid shifts between humid and semi-arid conditions drive high-frequency vegetation dynamics that are difficult to capture with retrospective or autoregressive methods alone. To address this gap, we developed a sequence-to-sequence ConvLSTM model that generates gridded, daily, 1-km LAI forecasts driven by daily weather data, explicitly producing the spatially continuous, prognostic LAI fields suitable for integration into coupled land-atmosphere and Earth system models. The model integrates a sequence of historical LAI maps together with observed meteorological data – including daily maximum temperature, minimum temperature, and precipitation – to predict LAI up to 30 days ahead across a continuous spatial domain. We focus model development and evaluation on the South-Central United States, a region encompassing diverse land cover and strong seasonal vegetation dynamics, serving as a proof-of-concept before scaling to the continental United States (CONUS). The remainder of this paper describes the data and study domain (Section 2), the model architecture and training configuration (Section 3), evaluation results and comparisons against baseline forecasts (Section 4), and a discussion of limitations and future directions (Section 5).

\section{Data and Preprocessing Methods}
\label{sec2}

This study integrates two primary data sources: satellite-derived LAI from MODIS and daily surface meteorological observations from Global Historical Climatology Network daily (GHCNd). Both datasets were preprocessed to spatially and temporally consistent 1-km daily grids prior to model training.

\subsection{Study Area}
\label{subsec2.1}

\begin{figure}[H]
\centering
\includegraphics[width=1\textwidth]{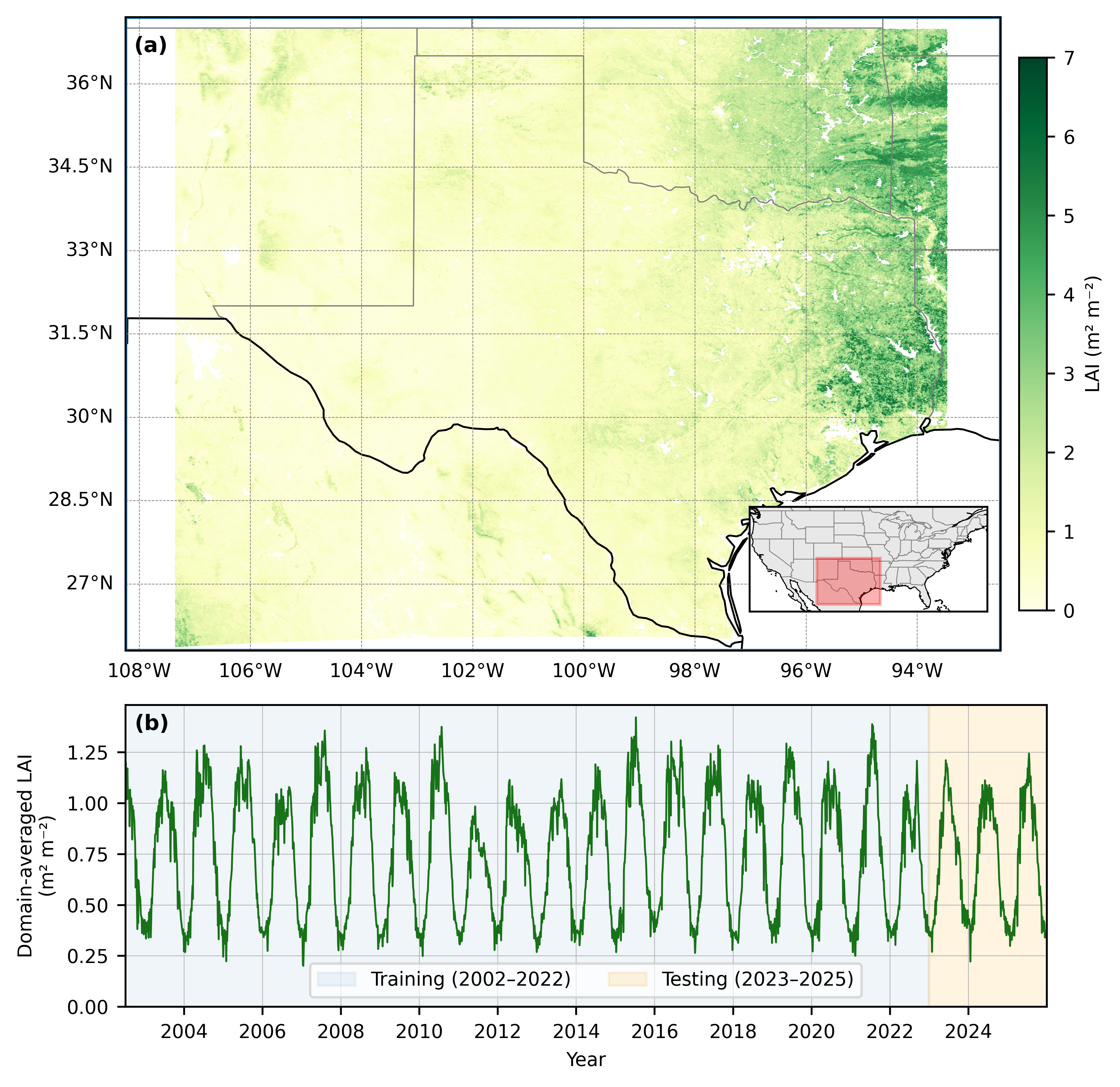}
\caption{Study domain and LAI characteristics. (a) Mean JJA LAI over the study domain for 2013. The red rectangle indicates the domain boundary. (b) Domain-averaged LAI time series from 2002 to 2025. Shaded regions indicate the training and validation period (2002–2022, blue), testing period (2023–2025, yellow). 
}\label{fig1}
\end{figure}

The study domain covers the South-Central United States, spanning 25.87°–37.17°N and 108.22°–92.59°W, encompassing portions of Texas, Oklahoma, Louisiana, Arkansas, New Mexico, and adjacent states. The region exhibits a pronounced west-to-east climate gradient, transitioning from semi-arid conditions in western Texas and New Mexico to humid subtropical conditions along the Gulf Coast and lower Mississippi Valley. This gradient drives strong spatial heterogeneity in vegetation cover and phenology, with LAI closely coupled to seasonal and interannual variability in temperature and precipitation. The western portion of the domain is additionally influenced by the North American Monsoon, which delivers a distinct pulse of summer precipitation that strongly modulates vegetation green up. This combination of climate diversity and strong atmosphere-vegetation coupling makes the domain well-suited for developing and evaluating a weather-driven LAI forecasting model. The spatial extent of the study domain and the LAI averaged over June, July and August (JJA) for 2013 are shown in Fig.~\ref{fig1}a (LAI data described in Section 2.2). Fig.~\ref{fig1}b presents the domain-averaged LAI time series from 2002 to 2025, illustrating both the strong seasonal cycle and interannual variability.

\subsection{MODIS LAI Data}
\label{subsec2.2}

LAI data were obtained from the MODIS Combined Terra and Aqua Leaf Area Index product (MCD15A3H, Version 6.1; \citealt{myneni2021}), distributed by the NASA Land Processes Distributed Active Archive Center. MCD15A3H provides 4-day composite LAI at 500 m resolution by selecting the best available observation from both Terra and Aqua sensors within each period.

Preprocessing involved quality control (QC), spatial aggregation, and temporal resampling to daily resolution. Pixel-level quality filtering was applied using two QC layers provided with the product: pixels flagged for significant cloud contamination in LAI/FPAR quality control layer (FparLai\_QC) were excluded, and only land-surface pixels with acceptable aggregated reflectance quality in extra quality control layer (FparExtra\_QC) were retained. Pixels failing either criterion were marked invalid in a binary LAI validity flag. 500 m pixels were then spatially aggregated to 1 km by averaging all contributing valid pixels within each target grid cell.

To align the 4-day LAI composites with daily meteorological forcing, LAI values were temporally resampled to daily resolution using a forward-filling scheme, in which each observation is held constant until the next valid composite becomes available. This approach preserves the integrity of the original retrievals without introducing artificial interpolation signals into model training. Days carrying forward a prior observation are identified in the LAI validity flag, distinguishing them from days with direct MODIS retrievals. The practical implication for model design is that the hidden state evolves daily with weather forcing while the LAI state is updated only when a real observation is available. The details are described in Section 3.

\subsection{Meteorological Forcing Data}
\label{subsec2.3}

Daily surface meteorological data were obtained from the Global Historical Climatology Network daily (GHCNd; \citealt{menne2012}), maintained by the National Centers for Environmental Information (NCEI). GHCNd provides quality-controlled daily summaries of temperature and precipitation from land surface stations across the study domain. Station density is spatially uneven, with greater concentration in populated lowlands and sparser coverage across arid and elevated terrain, which can introduce spatial bias in the interpolated fields, particularly in data-sparse regions.

\subsubsection{Station Network and Spatial Coverage}
\label{subsec2.3.1}

\begin{figure}[H]
\centering
\includegraphics[width=1\textwidth]{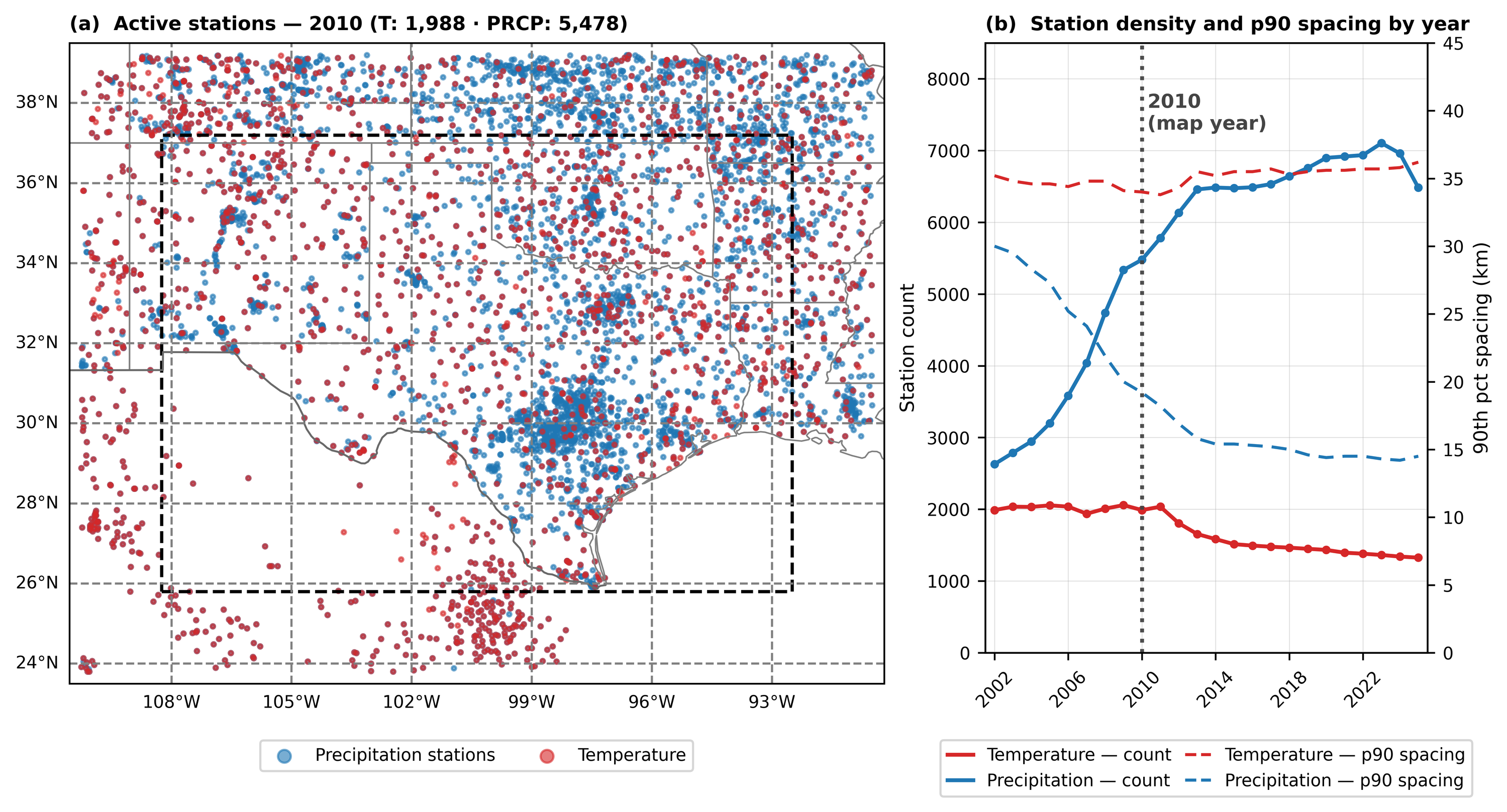}
\caption{(a) GHCNd station locations with records during 2010, representative of mid-period network density. Target domain shown as the dashed rectangle, with a station buffer used during interpolation. (b) Annual station count (solid lines, left axis) and inter-station distance at the 90th percentile (dashed lines, right axis) for temperature and precipitation stations, 2002–2025. Dotted vertical line marks the year shown in (a). 
}\label{fig2}
\end{figure}

The station networks used for interpolation varied in size throughout the study period. Temperature records were available from approximately 2,000 stations in 2002, declining to 1,325 by 2025, while precipitation records expanded nearly threefold, from 2,600 to 7,100 stations over the same period (Fig.\ref{fig2}).

Temperature station spacing remained consistent throughout the study period, with a mean inter-station distance of approximately 19 km and a 90th percentile of 35 km (Fig.\ref{fig2}b). Precipitation station spacing decreased substantially as the network expanded, with the mean inter-station distance declining from 17 km in 2002 to 6 km by 2025, and the 90th percentile decreasing from 30 km to 15 km over the same period. Despite adequate mean network density, both networks show spatial clustering toward populated areas, with sparser coverage over the arid western portions of the domain. Maximum observed inter-station gaps reached 275 km for temperature, motivating the use of a 2° buffer zone around the target domain to ensure adequate station coverage at the edges.

\subsubsection{Spatial Interpolation to Model Grid}
\label{subsec2.3.2}

Meteorological station observations were spatially interpolated to the 1 km model grid using variable-specific methods. For temperature, Barnes analysis \citep{barnes1964} was applied, a Gaussian distance-weighting scheme designed for interpolating irregularly spaced observations onto a regular grid. Temperature varies smoothly and continuously in space, making Barnes analysis physically appropriate. In this scheme, each grid point value is estimated as a weighted average of nearby station observations, with the weight decaying with distance as:

\begin{equation}
w(r) = \exp\left(-\frac{r^2}{4\kappa}\right)
\end{equation}

where $r$ is the distance from the station to the grid point and $\kappa$ is the smoothing parameter controlling the spatial influence. We set $\kappa$ following \citet{Koch1983} as 
$\kappa = (1.33\bar{d})^2 / 4$, where $\bar{d}$ is the mean inter-station spacing 
(${\sim}19$ km), yielding $\kappa \approx 160$ km$^2$. Two iterations were applied, with the second pass correcting residuals between station observations and the first-pass interpolated values to improve the fit to the original observations.

For precipitation, inverse distance weighting (IDW) was applied. Unlike Barnes analysis, IDW makes no assumptions about the spatial continuity or smoothness of the interpolated field --- each grid point value is estimated as a weighted average of nearby station observations, with weights decaying as $1/r^p$ where $r$ is the distance and $p$ is the power parameter. A maximum search radius of 50 km and a minimum of 3 stations were required for interpolation.

The nominal 1 km model of grid spacing does not reflect the true spatial resolution of the interpolated meteorological fields, which is governed by the station network density. Following \citet{barnes1964}, the effective resolution of the interpolated temperature field is approximately $2\bar{d} \approx 38$ km, corresponding to the minimum wavelength in the Fourier decomposition of the spatial field, given the mean inter-station spacing. Values at finer scales are mathematically interpolated with no gain in physical detail. To reduce introducing apparent spatial detail unsupported by the station network, temperature and precipitation fields were first interpolated to intermediate resolutions (20 km and 10 km) using Barnes analysis and IDW respectively, then bilinearly resampled to the 1 km model grid to match the LAI input resolution. No DEM-based lapse rate correction was applied to the temperature field, as the study domain is predominantly low-to-moderate relief terrain, and the correction would introduce additional uncertainty in areas of complex terrain where station density is lower.

\section{Spatiotemporal ConvLSTM Framework for LAI Forecasting}
\label{sec3}
The proposed forecasting framework utilizes a sequence-to-sequence ConvLSTM architecture to map multi-dimensional atmospheric and vegetation time series to future LAI states. By integrating spatial convolutions within recurrent cells, the model preserves the spatial information of gridded meteorological forcing while capturing the time-lagged response inherent in vegetation phenology. This section details the spatiotemporal configuration of the model inputs, the model architecture, sampling strategies, and the optimization procedures used for training.

\subsection{Spatiotemporal Input and Output Configuration}
\label{subsec3.1}

We developed the model to forecast daily LAI sequences up to two horizons: 14 days and 30 days. In each setting, the model utilizes a historical time series of length T (where T = 14 or 30) to generate a corresponding prognostic LAI sequence. A sequence-to-sequence formulation was adopted rather than training separate models for individual lead times. The primary objective is to produce continuous forecasts spanning subseasonal — from two weeks to two months in the future — for which a per-lead-time approach would require an impractically large number of independent models. Furthermore, because MODIS LAI is retrieved at a 4-day composite frequency and day-to-day LAI changes are small, a sequence-based architecture that shares temporal representations across lead times is both more efficient and better matching the slow varying nature of vegetation dynamics.

The selection of input variables was informed by the two-way feedbacks between LAI and atmospheric variables discussed in Section 1, and finalized through hyperparameter optimization \citep{bergstra2012random} on a randomly sampled subset of 8000 spatiotemporal patches with a fixed random seed for reproducibility (Section 3.5). Predictors are categorized as dynamic (time-varying) and static (time-invariant), and are summarized in Table~\ref{tab:inputs}. Meteorological forcing includes daily maximum and minimum temperature and precipitation, interpolated from GHCNd station observations (Section 2.2). A 14-day cumulative precipitation sum is additionally included to capture antecedent moisture conditions, as vegetation LAI responds to accumulated water availability over multi-day periods. Validity flags are provided for both the temperature and precipitation fields, marking grid cells where station coverage was insufficient during spatial interpolation.

\begin{table}[H]
\small
\centering
\caption{Summary of model input variables. Dynamic predictors vary 
across both time and space; static predictors are time-invariant.}
\label{tab:inputs}
\begin{tabular}{lllc}
\hline
\textbf{Category} & \textbf{Variable} & 
\textbf{Source} & \textbf{Channels} \\
\hline
\multicolumn{4}{l}{\textit{Dynamic Predictors (time-varying)}} \\
\hline
Meteorological & Daily max. temperature & 
GHCNd (interpolated) & 1 \\
Meteorological & Daily min. temperature & 
GHCNd (interpolated) & 1 \\
Meteorological & Daily precipitation & 
GHCNd (interpolated) & 1 \\
Meteorological & 14-day cumulative precipitation & 
GHCNd (interpolated) & 1 \\
Meteorological & Temperature validity flag & 
Derived (Section~\ref{subsec2.3}) & 1 \\
Meteorological & Precipitation validity flag & 
Derived (Section~\ref{subsec2.3}) & 1 \\
Vegetation & LAI & 
MODIS MCD15A3H & 1 \\
Vegetation & LAI validity flag & 
Derived (Section~\ref{subsec2.2}) & 1 \\
Land cover & Plant functional type (one-hot) & 
MODIS MCD12Q1 & 6 \\
\hline
\multicolumn{4}{l}{\textit{Static Predictors (time-invariant)}} \\
\hline
Temporal & Day of year (sinusoidal) & — & 2 \\
Spatial & Latitude & — & 1 \\
Spatial & Longitude & — & 1 \\
\hline
\textbf{Total} & & & \textbf{17}\\
\hline
\end{tabular}
\end{table}

Annual plant function type (PFT) derived from MODIS MCD12Q1 \citep{Friedl2022} are grouped into 6 classes as evergreen trees, deciduous trees, shrubs, grass, cereal croplands and broadleaf croplands. PFT, a proxy for vegetation parameters used to drive climate models \citep{Harper2023}, provides a categorical prior that constrains the expected phenological trajectory based specific vegetation functional traits.

Day of year is encoded using sine and cosine transformations, 
$\sin(2\pi \cdot \text{DOY}/365)$ and $\cos(2\pi \cdot \text{DOY}/365)$, preserving its cyclical continuity such that late December remains 
proximate to early January in the feature space. Latitude and longitude are included as static spatial context, providing geographic information 
that would otherwise be lost under the spatial subsampling strategy (Section~\ref{subsec3.3}).

The model output is a T-day sequence of predicted LAI values at 1 km resolution. All continuous input variables are z-score normalized prior to training.

\subsection{Model Architecture}
\label{subsec3.2}

The model is based on the ConvLSTM architecture \citep{shi2015}, which replaces linear operations in standard LSTM gates with 2D convolutions. This modification enables the network to capture spatiotemporal dependencies jointly. At each timestep the input gate $\mathbf{i}_t$,
forget gate $\mathbf{f}_t$, cell candidate $\mathbf{g}_t$, and output gate $\mathbf{o}_t$ are computed as:

\begin{align}
    \mathbf{i}_t &= \sigma\!\left(\mathbf{W}_{xi} * \mathbf{X}_t + \mathbf{W}_{hi} * \mathbf{H}_{t-1} + \mathbf{b}_i\right) \\
    \mathbf{f}_t &= \sigma\!\left(\mathbf{W}_{xf} * \mathbf{X}_t + \mathbf{W}_{hf} * \mathbf{H}_{t-1} + \mathbf{b}_f\right) \\
    \mathbf{g}_t &= \tanh\!\left(\mathbf{W}_{xg} * \mathbf{X}_t + \mathbf{W}_{hg} * \mathbf{H}_{t-1} + \mathbf{b}_g\right) \\
    \mathbf{o}_t &= \sigma\!\left(\mathbf{W}_{xo} * \mathbf{X}_t + \mathbf{W}_{ho} * \mathbf{H}_{t-1} + \mathbf{b}_o\right) \\
    \mathbf{C}_t &= \mathbf{f}_t \odot \mathbf{C}_{t-1} + \mathbf{i}_t \odot \mathbf{g}_t \\
    \mathbf{H}_t &= \mathbf{o}_t \odot \tanh(\mathbf{C}_t)
\end{align}

where $\mathbf{X}_t \in \mathbb{R}^{C \times H \times W}$ is the input tensor at timestep $t$, $*$ denotes 2D convolution, $\odot$ element-wise
multiplication, and $\sigma$ the sigmoid function. $\mathbf{W}$ and $\mathbf{b}$ are learnable convolutional weights and biases.

The model follows a sequence-to-sequence encoder-decoder design. During the encoding phase, the model processes the input sequence at each timestep, updating the hidden and cell states $(\mathbf{H}_t, \mathbf{C}_t)$ to condense the spatiotemporal history of the atmospheric and vegetation variables. The final states are then transferred to the decoder as its initial conditions. The decoder generates the output sequence autoregressively, producing a predicted LAI map at each lead time $\tau$:

\begin{equation}
    \hat{\mathbf{Y}}_\tau = f_\text{dec}\!\left(
        \hat{\mathbf{Y}}_{\tau-1},\, 
        \mathbf{H}^{(L)}_{\tau-1},\, 
        \mathbf{C}^{(L)}_{\tau-1}
    \right), \quad \tau = 1, \dots, T_\text{out}
\end{equation}
where $\hat{\mathbf{Y}}_0 = \mathbf{Y}_{T_\text{in}}$ is the last observed LAI map and superscript $(L)$ denotes the final encoder layer. Because each prediction depends on all prior predictions, errors can accumulate over longer horizons as a common limitation of autoregressive architectures.

We adopt a two-layer stacked ConvLSTM configuration. This stacking allows the second layer to extract deeper temporal abstractions from the hidden state representations produced by the first layer. Both layers utilize a hidden dimensionality of 16 channels and 3×3 convolutional kernels. The 3×3 kernel captures spatial context from the eight immediate neighbors, corresponding to a 3 km × 3 km receptive field at our 1 km resolution. 

\begin{figure}[H]
\centering
\includegraphics[width=1\textwidth]{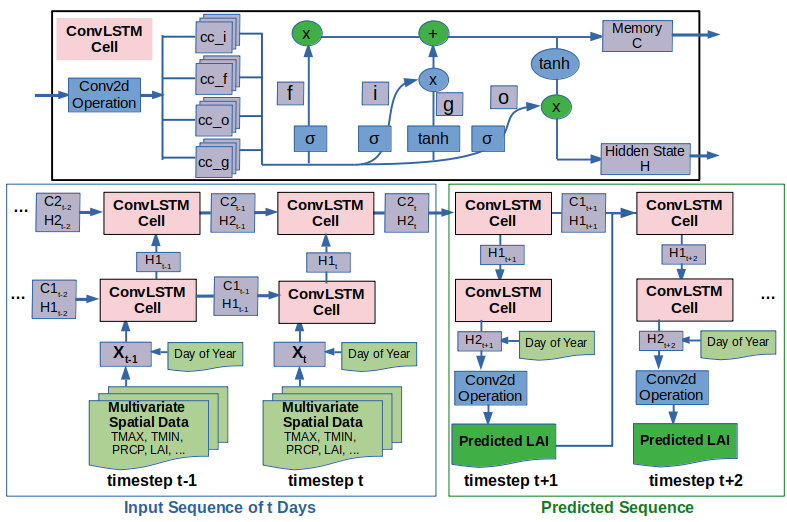}
\caption{(\textit{Top}) Internal gate structure of a single ConvLSTM cell, showing the input ($\mathbf{i}$), forget ($\mathbf{f}$), output ($\mathbf{o}$), and cell candidate ($\mathbf{g}$) gates. (\textit{Bottom}) Unrolled encoder-decoder sequence: the encoder processes $t$ days of multivariate spatial inputs augmented with day-of-year encoding; the decoder autoregressively generates the predicted LAI sequence, conditioned on the final encoder states.
}\label{fig_architecture}
\end{figure}

To mitigate overfitting and improve generalization, dropout with a rate of 0.2 is applied to the hidden state of the first encoder layer before it is passed to the second layer. At each timestep, the four ConvLSTM gates – input gate i, forget gate f, output gate o, and cell candidate – control the flow of information, determining which new features are written to memory and which past memory should be retained to drive the prognostic output. The overall architecture is illustrated in Fig.~\ref{fig_architecture}

\subsection{Spatial Subsampling Strategy}
\label{subsec3.3}
Training the LAI forecasting model on the full 1 km resolution study domain is computationally intensive, with more than 1.7 million pixels. The scale would grow to approximately 15.3 million pixels at CONUS extent. Training on all available pixels is also methodologically inadvisable, as spatially adjacent pixels share highly correlated inputs and targets, contributing redundant information to parameter updates and increasing the risk of overfitting. 

To address this, the domain was partitioned into non-overlapping 100 × 100-pixel patches for spatial sampling. For the final models, all patches across the domain were used for training, ensuring complete spatial coverage while reducing spatial redundancy between adjacent samples. Temporally, for each spatial patch, a stride equal to the input sequence length was applied so that consecutive temporal samples did not overlap in time, further reducing redundancy. This procedure yielded approximately 72,000 and 38,000 spatiotemporal training patches for the 14-day and 30-day configurations, respectively. For the hyperparameter experiments described in Section 3.4, a randomly sampled subset of 8,000 patches was drawn from this pool to enable efficient exploration of the configuration space.

\subsection{Loss Function and Training}
\label{subsec3.4}
The model is trained by minimizing the mean squared error (MSE) between predicted and observed LAI, computed on all pixels except those flagged as invalid (flag=0), such as water bodies and cloud-contaminated retrievals. Because MODIS LAI is available only as a 4-day composite products, restricting the loss function to true-retrieval pixels (flag=1) would leave the most days in the 14- and 30-day prediction sequences without direct supervision. In preliminary experiments, this sparse-supervision approach showed slightly weaker performance, with larger bias in certain regions and seasons. We nonetheless include forward-filled pixels (flag=2) in the loss computation, while retaining the validity flag as an input channel (Section 3.1), so the model can distinguish genuine retrievals from interpolated values. This enables a denser training signal across the full sequence length while preserving the model's ability to learn from the timing of true observational updates. The loss is applied uniformly across all predicted timesteps and valid spatial pixels, with no seasonal or spatial weighting. Formally, the masked MSE loss is defined as:

\begin{equation}
\mathcal{L} = \frac{1}{\sum_{t,i,j} \mathbf{1}[f_{t,i,j} \in \{1,2\}]} 
\sum_{t,i,j} \mathbf{1}[f_{t,i,j} \in \{1,2\}] \cdot 
\left(\hat{y}_{t,i,j} - y_{t,i,j}\right)^2
\end{equation}

where $\hat{y}_{t,i,j}$ and $y_{t,i,j}$ denote the predicted and observed z-scored LAI at timestep $t$ and spatial location $(i,j)$; $f_{t,i,j} \in \{0, 1, 2\}$ is the LAI validity flag; and $\mathbf{1}[\cdot]$ is the indicator function that equals 1 when $f_{t,i,j} \in \{1,2\}$ (valid MODIS retrieval or interpolated observation) and 0 otherwise.

The primary model targets a 30-day forecast horizon and it was trained using the Adam optimizer with an initial learning rate of 0.0003 and a batch size of 64. The learning rate was reduced by a factor of 0.5 when the validation loss showed no improvement for 20 consecutive epochs, with a minimum learning rate of 1e-6. This learning rate schedule allows the optimizer to make finer weight adjustments as training progresses. Early stopping was applied based on validation loss with a patience of 30 epochs and a minimum improvement threshold of 0.0001. The training data were split chronologically: samples from 2002 through 2021 were used for training, 2022 was reserved for validation, and 2023–2025 was  held out for independent evaluation. The 30-day model was trained on samples constructed with a 30-day stride, approximately 33,000 valid training samples.

\subsection{Selection of Kernel Size and Input Sequence Length}
\label{subsec3.5}

Selecting model hyperparameters (e.g., learning rate, batch size, hidden dimensionality, and number of filters) required balancing forecast accuracy against model complexity, computational cost, and the risk of overfitting. We conducted a series of exploratory experiments on a randomly sampled subset of 8,000 spatiotemporal patches (fixed random seed for reproducibility), with each configuration trained for 20 epochs and repeated across 10 independent runs to account for initialization variability. While additional hyperparameters (e.g., dropout rate, learning rate schedule, number of stacked layers) were also tuned during model development, we focus this section on kernel size and input sequence length — two hyperparameters that directly reflect the spatial and temporal correlation structure of LAI changes.

\begin{figure}[H]
\centering
\includegraphics[width=1\textwidth]{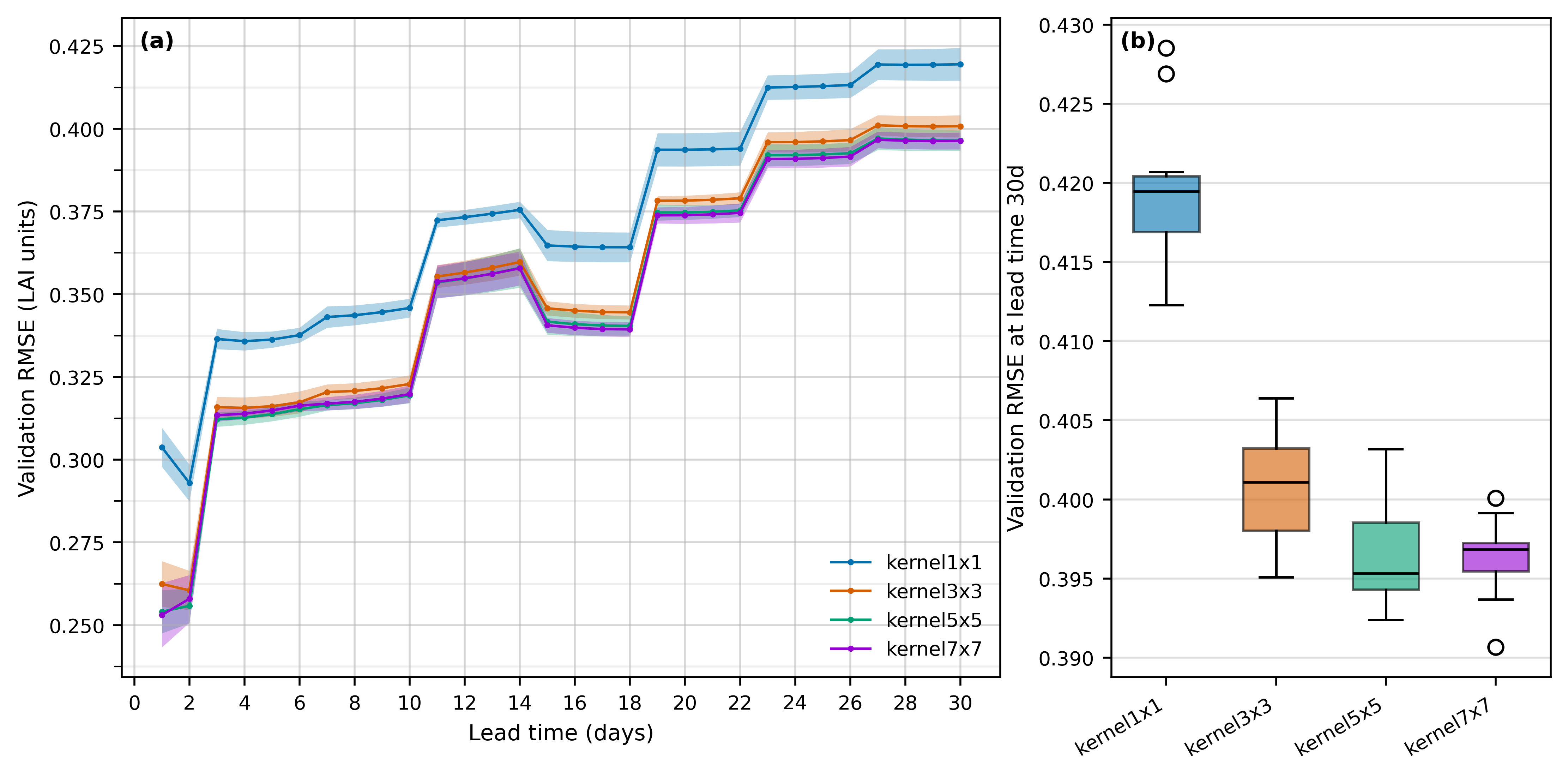}
\caption{Exploratory experiments for hyperparameter tuning of kernel size. (a) Validation RMSE as a function of lead time for each kernel size. Shaded regions show variation across 10 repetitions. (b) Box plot of validation RMSE at a 30-day lead time.  
}\label{fig_kernel}
\end{figure}

\begin{figure}[H]
\centering
\includegraphics[width=1\textwidth]{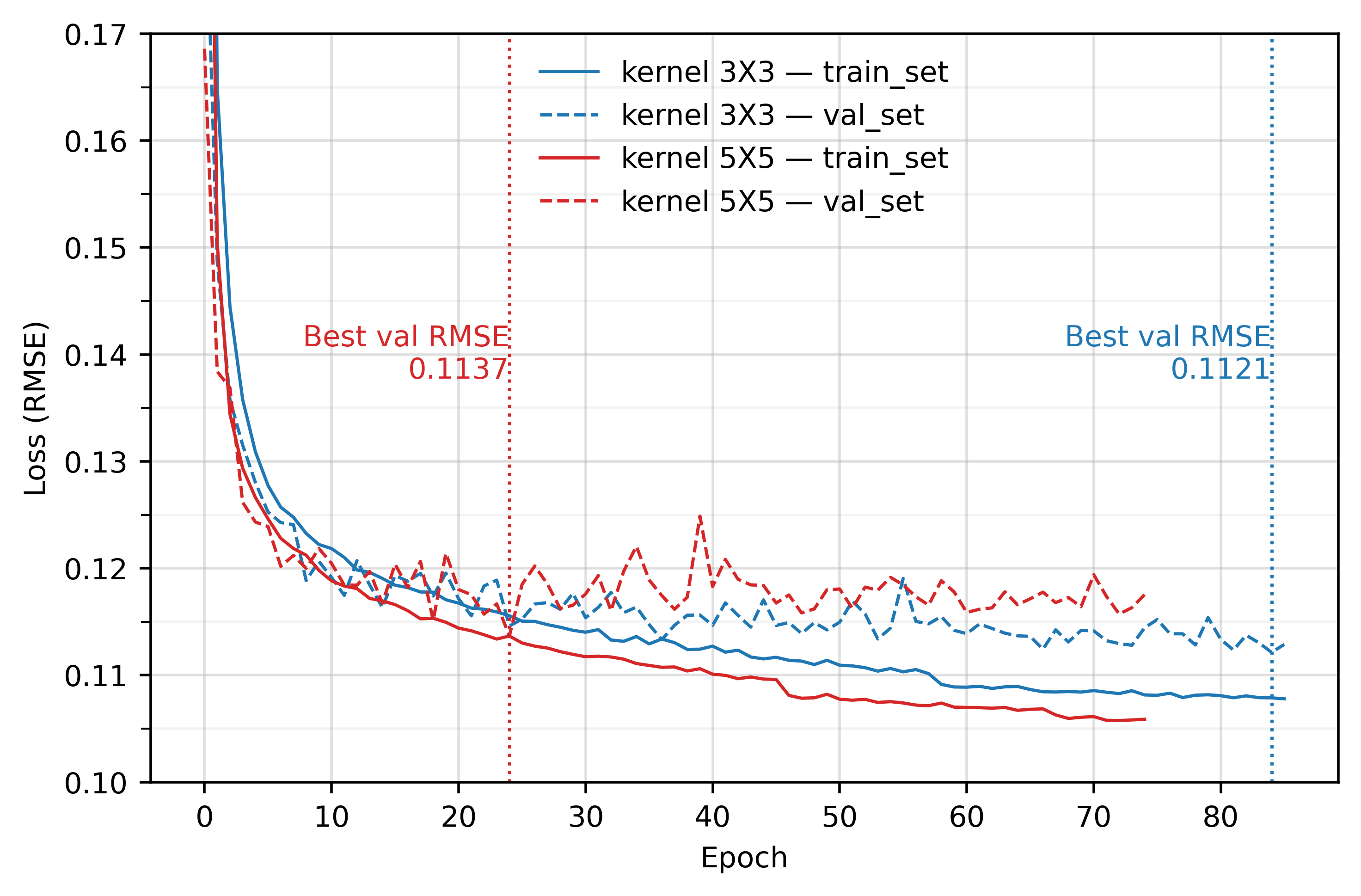}
\caption{Training and validation loss during full-scale model training for 3$\times$3 and 5$\times$5 kernel configurations. Solid and dashed lines show training and validation RMSE, respectively, as a function of epoch; blue indicates the 3$\times$3 kernel and red indicates the 5$\times$5 kernel. Vertical dotted lines mark the epoch of best validation RMSE for each configuration.}
\label{fig_kernel_training}
\end{figure}

Figure~\ref{fig_kernel} presents validation RMSE across kernel sizes of 1$\times$1, 3$\times$3, 5$\times$5, and 7$\times$7, with all other settings held fixed as described in Section 3.2. Kernel size controls the spatial extent of neighboring information incorporated at each timestep. Within this reduced-data exploratory setting, the 1$\times$1 (pixel-wise) kernel yielded the highest RMSE, while the 5$\times$5 kernel achieved slightly lower validation RMSE than the 3$\times$3 kernel (mean RMSE of 0.3965 vs. 0.4008 at a 30-day lead time). The 7$\times$7 showed marginally lower RMSE still, but this gain was not proportional to its substantially higher parameter count relative to the other configurations (Table~\ref{tab2}). Based on these results, the 3$\times$3 and 5$\times$5 kernels were selected as candidates for full-scale training and evaluation on the independent test set. During full-scale training, the 5$\times$5 configuration reached its lowest validation RMSE (0.1137) within 25 epochs before validation loss became increasingly noisy and trended upward, while training loss continued to decrease --- a signature of overfitting (Fig.~\ref{fig_kernel_training}). The 3$\times$3 configuration maintained a more stable validation loss throughout training and achieved a lower best validation RMSE (0.1121). This pattern was also reflected in spatial, seasonal, and PFT-stratified error metrics, all of which favored the 3$\times$3 configuration. We therefore selected the 3$\times$3 kernel as the final model configuration.

\begin{table}[H]
\centering
\caption{The size of trainable parameters for each model configuration.}
\label{tab2}
\begin{tabular}{lc}
\toprule
Configuration & Trainable parameters \\
\midrule
Kernel size 1$\times$1 & 7,635 \\
Kernel size 3$\times$3 & 66,515 \\
Kernel size 5$\times$5 & 184,275 \\
Kernel size 7$\times$7 & 360,915 \\
\bottomrule
\end{tabular}
\end{table}

\begin{figure}[H]
\centering
\includegraphics[width=1\textwidth]{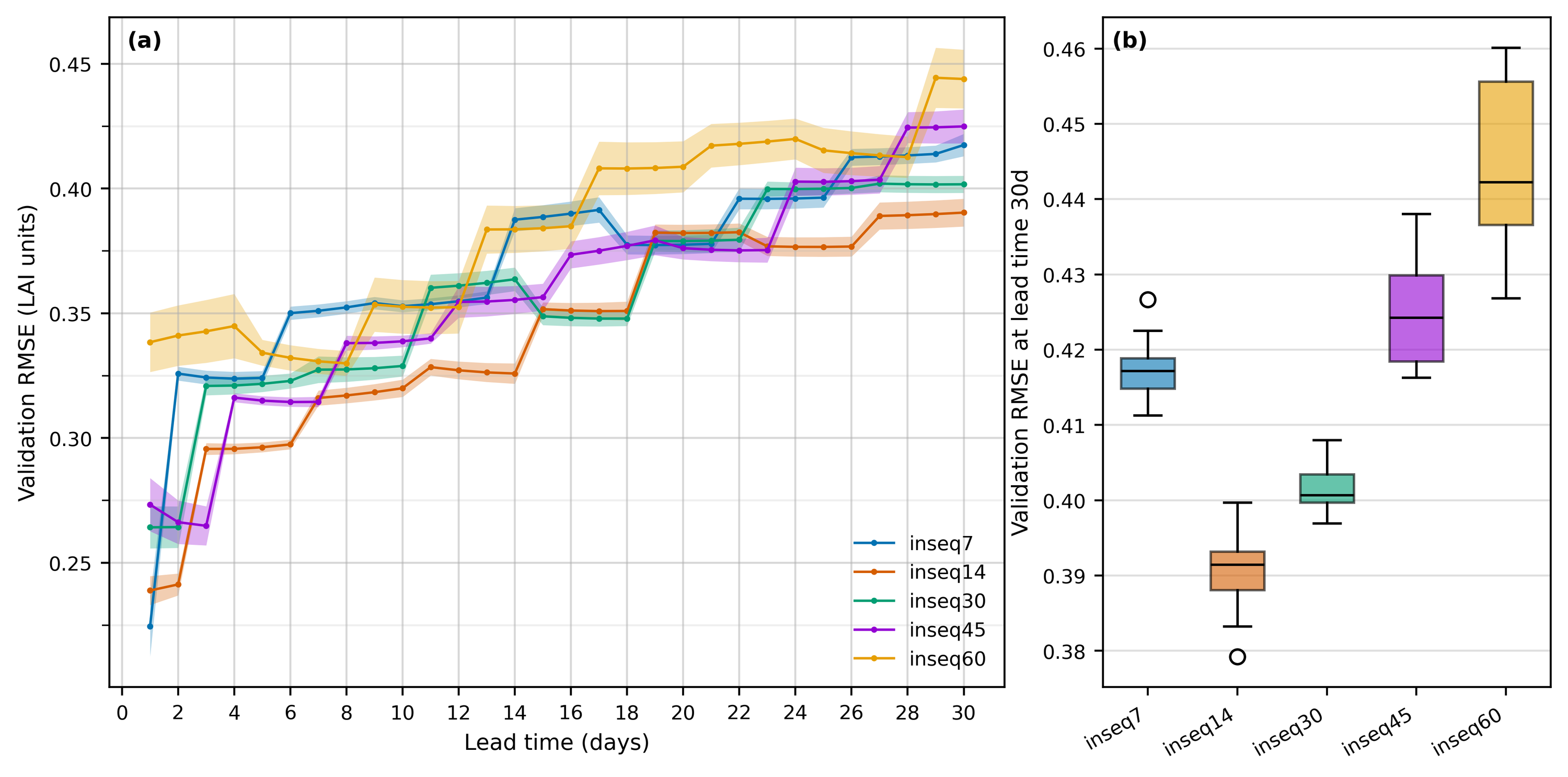}
\caption{Exploratory experiments for hyperparameter tuning of input sequence length. (a) Validation RMSE as a function of lead time for each length. Shaded regions show variation across 10 repetitions. (b) Box plot of validation RMSE at a 30-day lead time.}
\label{fig_seqlen}
\end{figure}

Figure~\ref{fig_seqlen} shows validation RMSE as a function of lead time for the input sequence length experiments, in which input sequences of 7, 14, 30, 45, and 60 days were tested with the output forecast length fixed at 30 days. Input sequences of 14 and 30 days achieved the lowest validation RMSE, with the 14-day input performing slightly better than the 30-day input. Longer input sequences did not yield further improvement, and the 60-day input sequence showed the worst performance, possibly reflecting a general limitation of recurrent architectures in effectively using very long input histories. Based on these results, two primary model configurations were developed with matched input and output sequence lengths of 14 and 30 days, respectively. While the experiments (Fig.~\ref{fig_seqlen}) indicate that a 14-day input alone achieves comparable skill for 30-day forecasts, we retained matched input and output lengths to ensure sufficient historical context across the full forecast horizon, which is also a common design choice in sequence-to-sequence forecasting.

\section{Model Evaluation}
\label{sec4}
Model performance is evaluated across two complementary aspects. First, overall forecast accuracy is quantified across varying lead times using standard machine learning metrics: root mean square error (RMSE), which is sensitive to large errors due to the squared penalty; mean absolute error (MAE), which provides an interpretable measure of average error magnitude; and the coefficient of determination ($R^2$), which quantifies the proportion of LAI variance explained by the model. The second aspect focuses on the biophysical interpretation of LAI forecasts, examining model performance across seasons, geographic distributions, and vegetation types. The pronounced seasonal variability and land cover dependence of LAI produce dynamic, spatially heterogeneous patterns that are challenging to capture, making domain-specific evaluation essential. Unless otherwise noted, RMSE, MAE, and $R^2$ are computed only on pixels with true MODIS retrievals (flag=1), excluding forward-filled pixels (flag=2), to ensure that reported skill reflects performance against real observations. Metrics are aggregated over all such valid pixels within the domain and averaged across the full testing period (2023--2025).

\subsection{Forecast Skill Across Lead Times}
\label{subsec4.1}

Forecast skill is additionally compared against a persistence forecast as a reference baseline, defined as:
$$LAI_{forecast}(t + \tau) = LAI_{obs}(t)$$
where $\tau$ represents the lead time in days. Persistence is a standard baseline in sub-seasonal to seasonal forecasting, representing a particularly meaningful benchmark for vegetation variables given their relatively slow temporal dynamics.

\begin{figure}[H]
\centering
\includegraphics[width=1\textwidth]{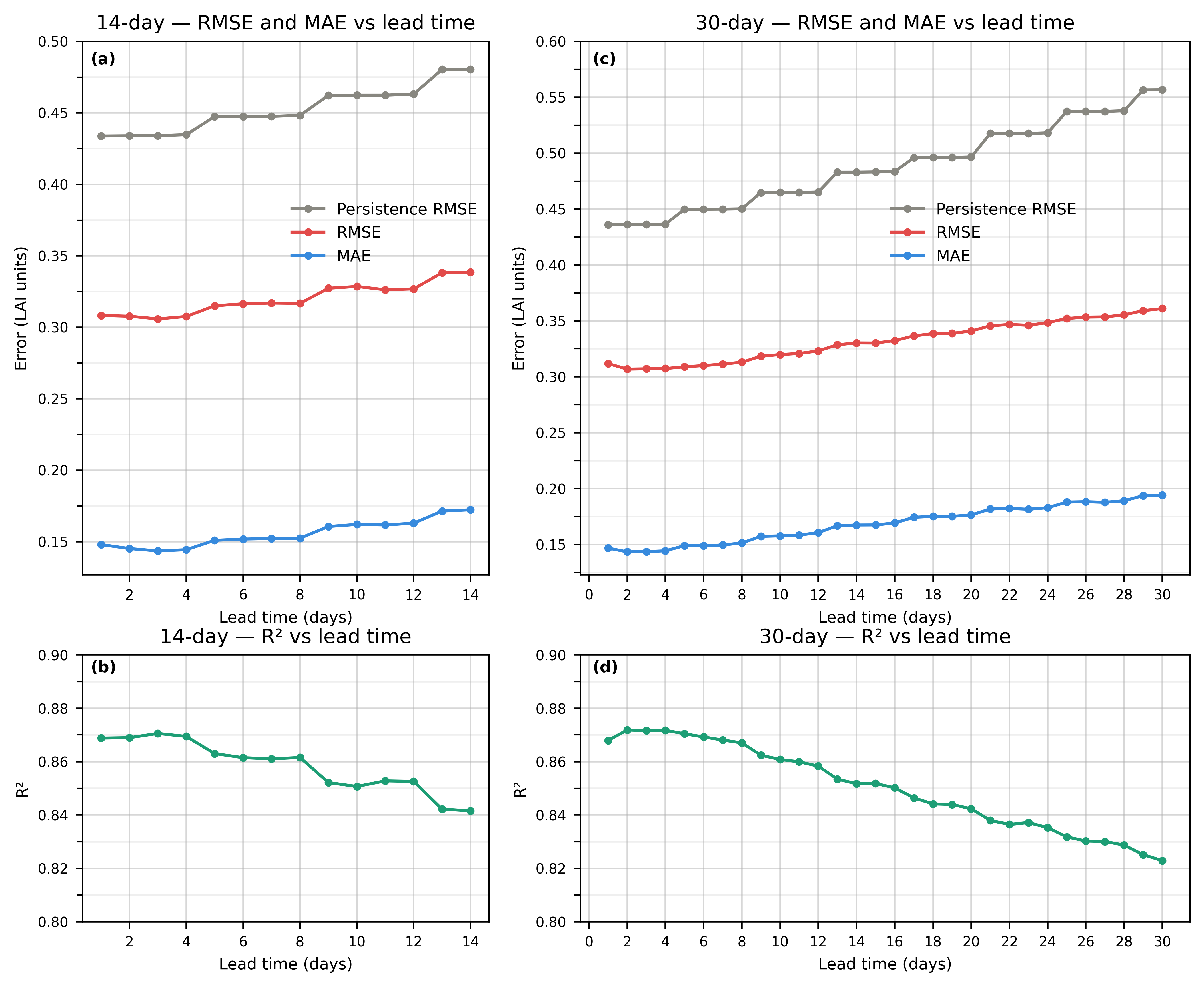}
\caption{Forecast skill across lead times for the 14-day and 30-day LAI forecasting models, evaluated over the independent testing period (2023--2025). (a) RMSE (red) and MAE (blue) for the 14-day model, compared against the persistence forecast RMSE (gray) as a reference baseline. (b) $R^2$ for the 14-day model across lead times 1--14 days. (c) Same as (a) but for the 30-day model across lead times 1--30 days. (d) Same as (b) but for the 30-day model. All metrics are computed pixel-wise over quality-controlled valid pixels within the study domain and spatially averaged across the evaluation period. Error units are LAI units ($\text{m}^2\,\text{m}^{-2}$).}
\label{fig:metrics_leadtime}
\end{figure}

Both the 14-day and 30-day models demonstrate consistent forecast skill across their respective lead times, with RMSE and MAE increasing gradually and $R^2$ remaining stable as the forecast horizon extends (Fig.~\ref{fig:metrics_leadtime}). For the 14-day model, RMSE increases from 0.31 to 0.34 (lead time 1--14 days), while $R^2$ remains high throughout (0.87--0.84), indicating that the model captures the dominant variance in LAI dynamics even at longer lead times. The 30-day model shows a similar pattern of gradual skill decrease, with RMSE increasing from 0.30 to 0.36 (lead time 1--30 days) and $R^2$ ranging from 0.87 to 0.82. Both models substantially outperform the persistence baseline across all lead times --- the 14-day model reduces RMSE by 30\% relative to persistence, while the 30-day model achieves a 35\% reduction --- demonstrating meaningful forecast skill beyond a constant-state assumption.

Domain-level metrics are comparable across the two configurations, suggesting that the model framework is robust for both sequence lengths. However, aggregated metrics alone may obscure differences in performance at specific locations, seasons, or vegetation types. These are examined in detail in subsequent sections.

\subsection{Spatial and Seasonal Distribution of Forecast Errors}
\label{subsec4.2}

LAI exhibits pronounced spatial heterogeneity driven by distinct vegetation types, geographic constraints, and localized weather conditions. While the domain-averaged metrics presented in Section 4.1 quantify overall model accuracy, they can mask regional variation in performance. Characterizing the spatial and seasonal distribution of forecast errors is therefore essential for identifying where the model framework underperforms, and provides diagnostic feedback to guide future expansion to the broader CONUS domain.

\subsubsection{Spatial Distribution of Forecast Errors}
\label{subsec4.2.1}

\begin{figure}[H]
\centering
\includegraphics[width=1\textwidth]{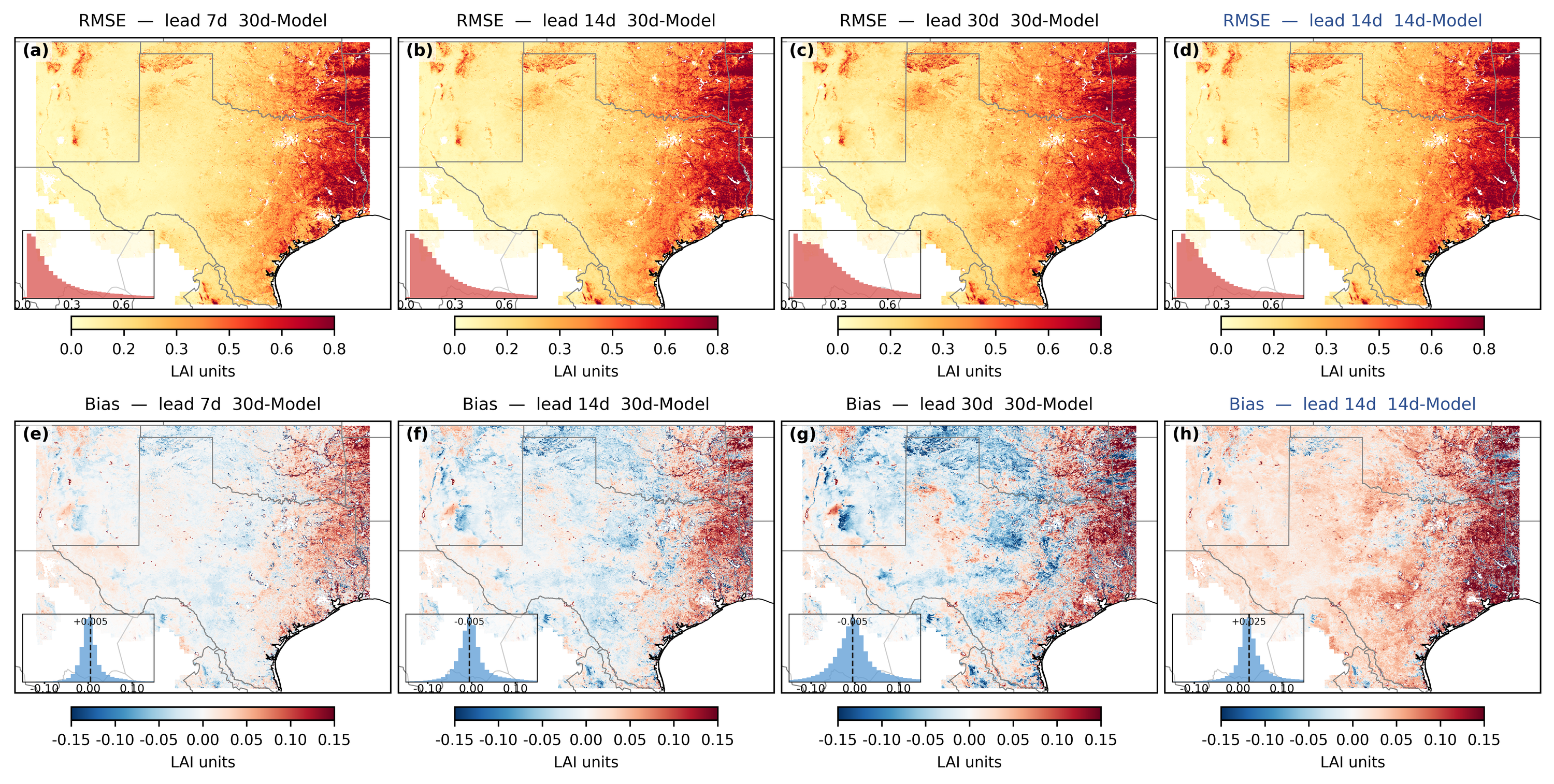}
\caption{Spatial distribution of pixel-wise RMSE (a--d) and bias (e--h) over the independent testing period (2023--2025). (a--c, e--g) 30-day model at lead times of 7, 14, and 30 days. (d, h) 14-day model at a 14-day lead time, for comparison. Accompanying histograms show the domain-wide frequency distribution of pixel-wise errors.}
\label{fig:spatial_compare}
\end{figure}

Figure~\ref{fig:spatial_compare} presents the spatial distribution of pixel-wise RMSE and bias at lead times of 7, 14, and 30 days from the 30-day model, alongside the 14-day lead time from the 14-day model for comparison. Accompanying histograms quantify the domain-wide distribution of these errors. Spatial patterns of RMSE are broadly similar between the two model configurations at the 14-day lead time, consistent with the comparable domain-averaged metrics reported in Section 4.1. Because RMSE scales with the magnitude of the predicted variable, regions with higher LAI values tend to exhibit larger absolute errors. The highest RMSE values are concentrated in the eastern portion of the domain, where Deciduous Broadleaf Forest supports LAI values substantially greater than those in the central and western regions. These elevated errors are therefore largely proportional to the local leaf area index rather than indicative of a systematic model deficiency.

While RMSE captures the magnitude of errors, bias maps reveal whether the model systematically over- or underestimates LAI spatially. A region with high RMSE but near-zero bias indicates that large errors occur but cancel across the evaluation period, whereas a region with low RMSE but notable bias indicates consistent directional error. The 30-day model demonstrates more spatially balanced bias than the 14-day model. Bias also increases progressively from the 7-day to the 30-day lead time, consistent with the seq2seq architecture, in which prediction errors can accumulate across successive forecast steps. Notably, the 14-day model exhibits an overestimation bias of approximately 0.025 LAI units relative to the 30-day model at the equivalent 14-day lead time --- despite being specifically optimized for shorter-horizon prediction, it shows greater systematic bias than the 30-day model does at that same horizon. This suggests that extended meteorological input context may play a more important role in LAI forecasting accuracy than forecast horizon length alone.

\subsubsection{Seasonal Bias Patterns}
\label{subsec4.2.2}

\begin{figure}[H]
\centering
\includegraphics[width=1\textwidth]{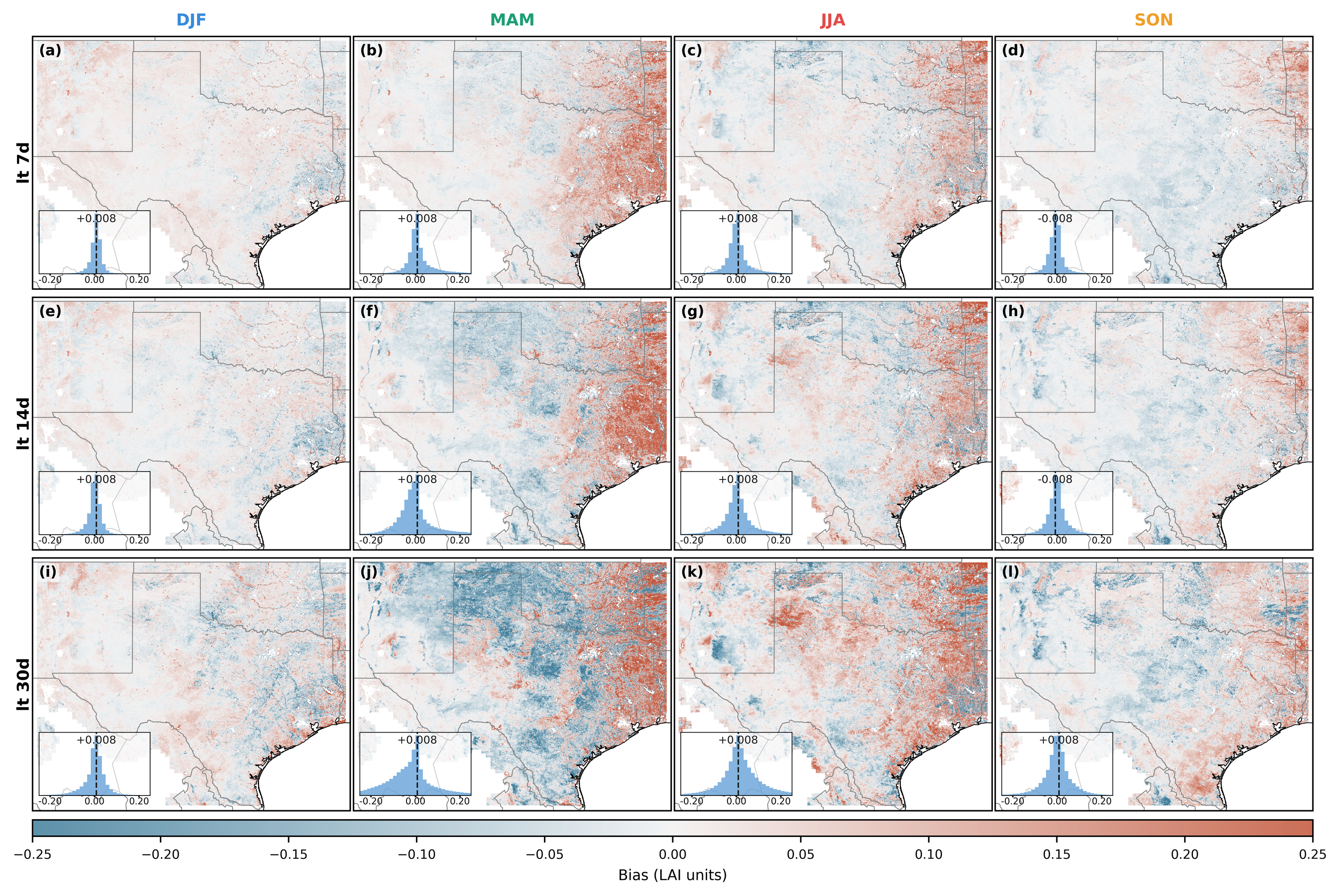}
\caption{Seasonal spatial bias maps from the 30-day model over the independent testing period (2023--2025). Each row corresponds to a lead time of 7, 14, or 30 days, and each column corresponds to a meteorological season (DJF, MAM, JJA, SON). Bias is defined as the mean signed error (predicted minus observed), computed pixel-wise and averaged across the evaluation period.}
\label{fig:seasonal_bias}
\end{figure}

\begin{figure}[H]
\centering
\includegraphics[width=1\textwidth]{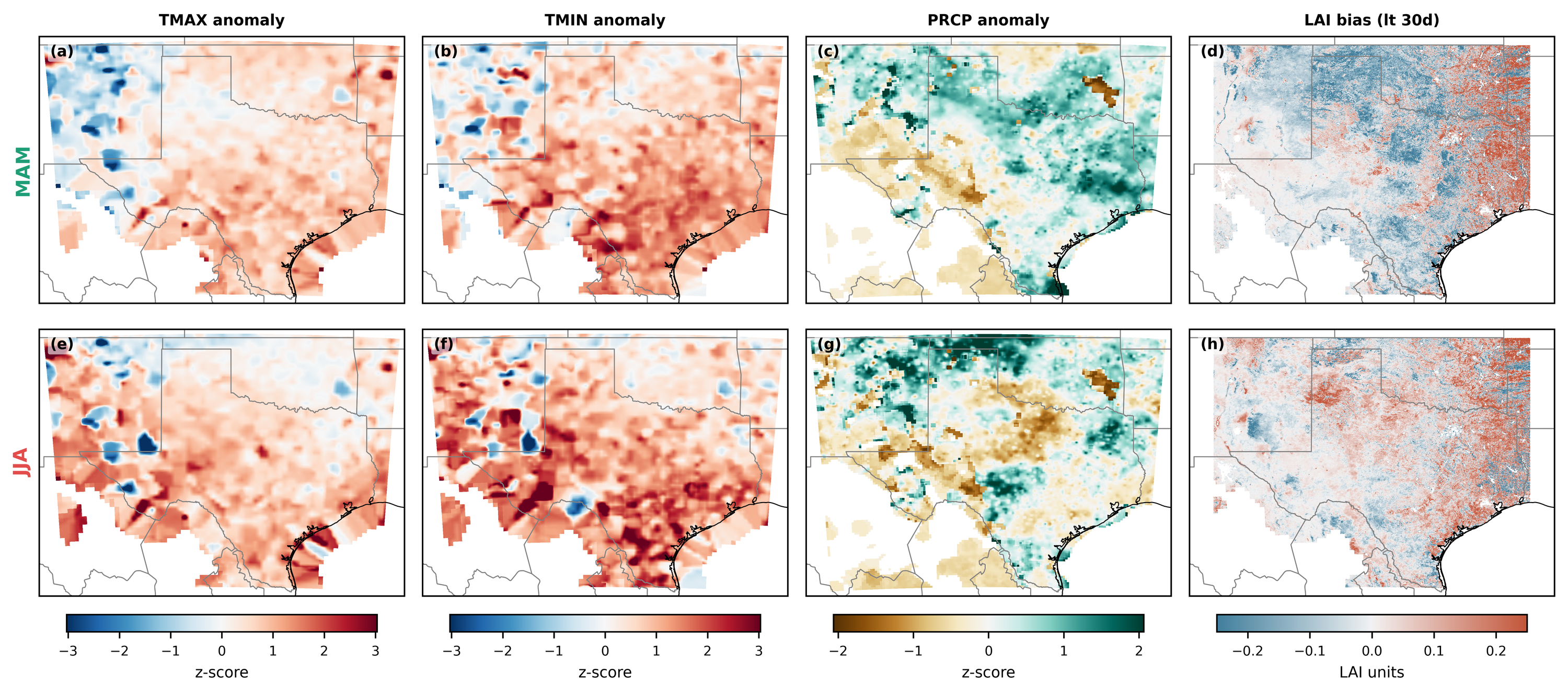}
\caption{Seasonal meteorological anomalies and corresponding 30-day-lead-time LAI bias over the central domain grassland region, for MAM (top row) and JJA (bottom row), 2023--2025. (a, e) TMAX anomaly (z-score) relative to the 2002--2022 climatology. (b, f) TMIN anomaly (z-score). (c, g) Precipitation anomaly (z-score). (d, h) LAI bias at a 30-day lead time.}
\label{fig:met_anomaly_bias}
\end{figure}

The seasonal spatial bias maps (Figure~\ref{fig:seasonal_bias}) show pixel-wise bias at lead times of 7, 14, and 30 days across the four meteorological seasons --- December--February (DJF), March--May (MAM), June--August (JJA), and September--November (SON). MAM and JJA exhibit the largest spatial bias, coinciding with the active growing season when LAI variability is greatest. Notably, the 30-day model underestimates LAI in the central domain during MAM --- a region predominantly covered by grassland --- and overestimates the same region during JJA. This reversal is consistent with the phenological cycle of grassland vegetation: during spring green-up, rapid LAI increases outpace the model's prediction, while during peak summer the model may anticipate continued growth that is instead constrained by drought stress or moisture limitation in the semi-arid central domain. This tendency is amplified at the 30-day lead time, where the model relies on meteorological conditions from 30 days prior and has no knowledge of subsequent weather evolution. This interpretation is supported by the seasonal meteorological anomalies over the same period (Figure~\ref{fig:met_anomaly_bias}): the central domain received above-normal precipitation during MAM, spatially coincident with the region of LAI underestimation, and below-normal precipitation during JJA, spatially coincident with the region of LAI overestimation. In contrast, TMAX and TMIN anomalies were positive across nearly the entire domain in both seasons, with comparatively little spatial structure distinguishing the central grassland from surrounding regions. The spatial pattern of forecast bias therefore more closely tracks the heterogeneous precipitation anomaly than the spatially uniform temperature anomaly, suggesting that precipitation variability, rather than temperature, is the primary driver of the bias in this grassland region --- consistent with the dominant role of precipitation in LAI variability reported by \citet{Lu1999}. DJF and SON show substantially lower and spatially uniform bias, reflecting the relatively low LAI variability during dormant and senescent periods.

\begin{figure}[H]
\centering
\includegraphics[width=1\textwidth]{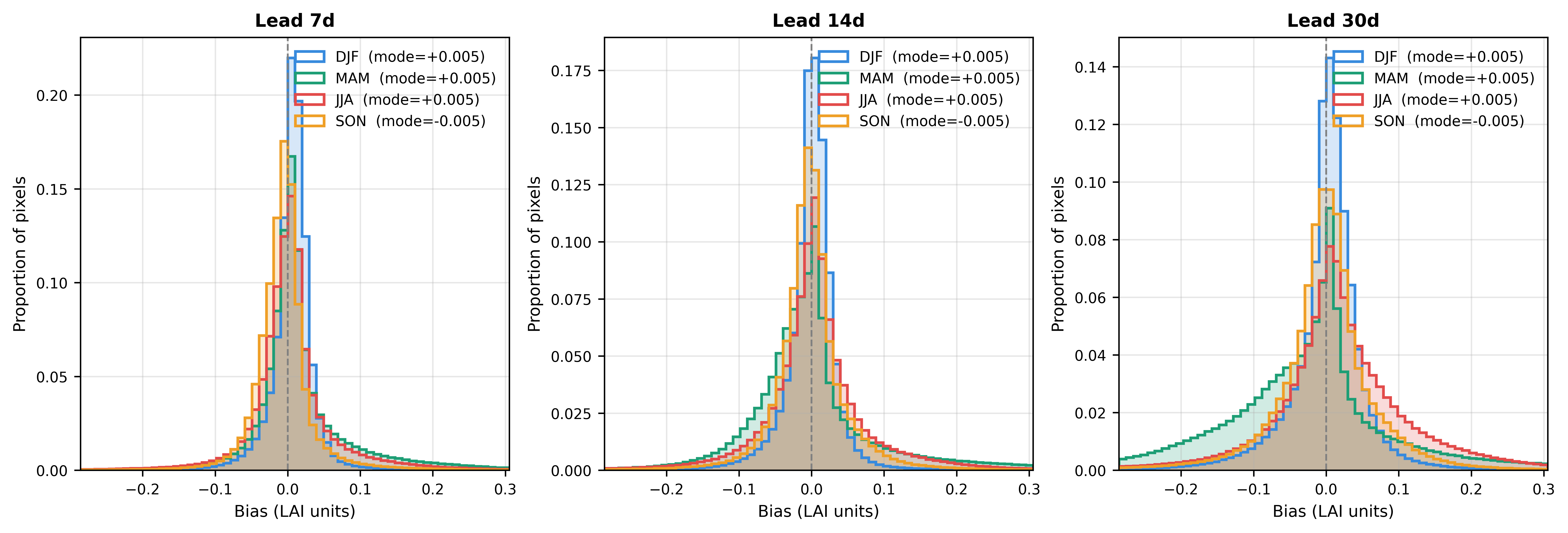}
\caption{Seasonal bias distributions from the 30-day model at lead times of 7 (a), 14 (b), and 30 days (c), over the independent testing period (2023--2025). Each panel shows the frequency distribution of pixel-wise bias across the domain for the four meteorological seasons (DJF, MAM, JJA, SON). The vertical dashed line at zero indicates unbiased prediction.}
\label{fig:seasonal_hist}
\end{figure}

Figure~\ref{fig:seasonal_hist} presents the corresponding bias distributions quantitatively for each season and lead time. The center of each distribution indicates the mean bias direction --- a distribution centered to the right of zero indicates systematic overestimation, while one centered to the left indicates underestimation. The width and shape of the distribution reflect the spread of errors: a narrow distribution indicates spatially consistent predictions, a wide distribution indicates high spatial variability in bias, and a skewed distribution indicates systematic directional errors across the domain. At lead times of 7 and 14 days, bias distributions are narrow and approximately centered near zero across all four seasons, demonstrating stable seasonal performance. At a lead time of 30 days, distributions are broader as expected for a longer forecast horizon, yet remain relatively concentrated near zero. MAM and JJA show wider and more skewed distributions compared to DJF and SON, confirming that the model captures the low-variability seasons more easily.

\begin{figure}[H]
\centering
\includegraphics[width=1\textwidth]{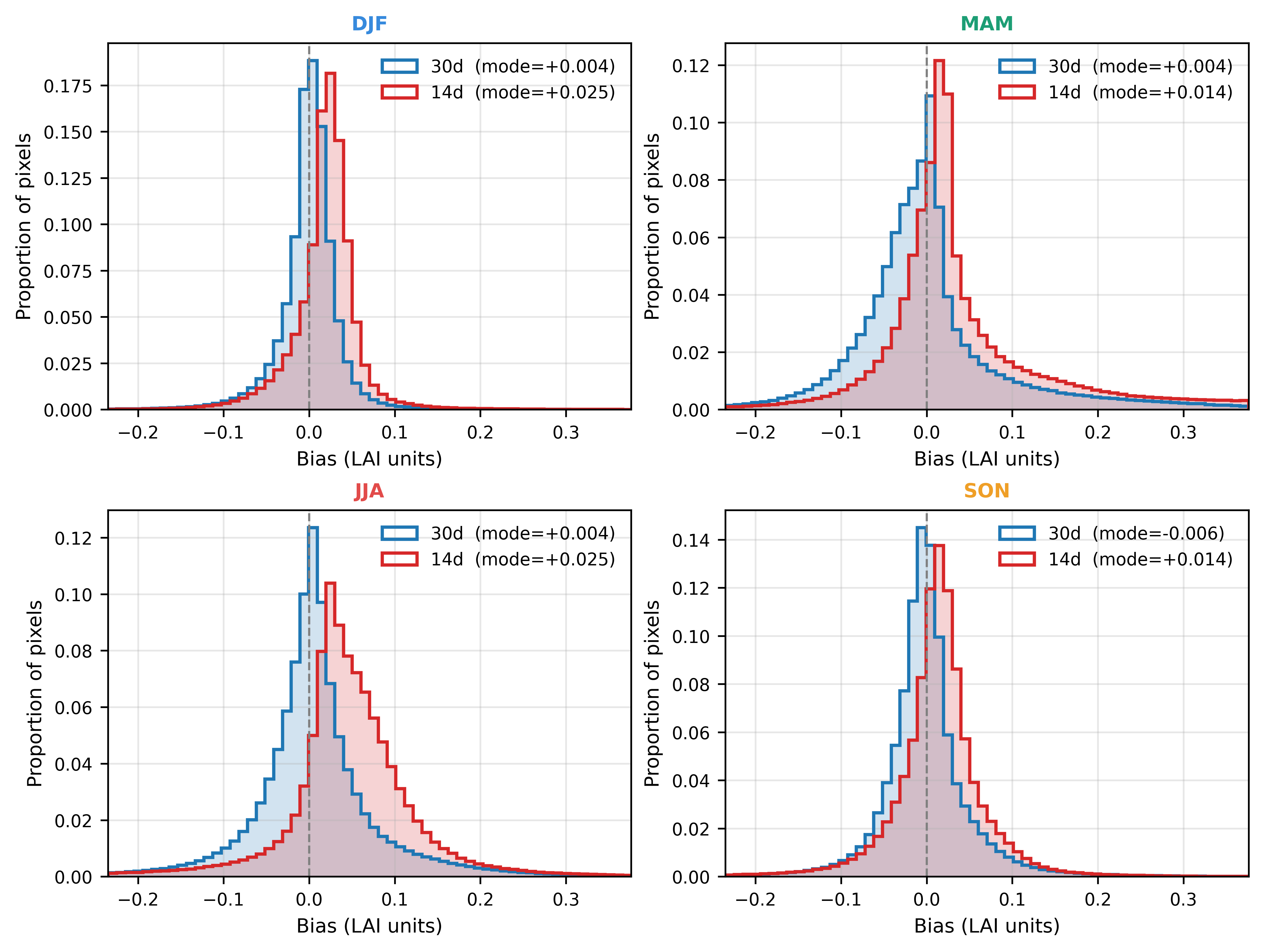}
\caption{Comparison of bias distributions at a 14-day lead time between the 14-day and 30-day models across the four meteorological seasons: (a) DJF, (b) MAM, (c) JJA, and (d) SON. Each panel shows the frequency distribution of pixel-wise bias for both model configurations over the independent testing period (2023--2025). The vertical dashed line at zero indicates unbiased prediction.}
\label{fig:seasonal_bias_dist_compare}
\end{figure}

Figure~\ref{fig:seasonal_bias_dist_compare} compares the bias distributions at a 14-day lead time from the 14-day and 30-day models across all four seasons. Both models exhibit bias values mainly within $\pm$0.2 LAI units and show similar distributions in DJF and SON. In MAM and JJA, the 30-day model produces a more zero-centered bias distribution than the 14-day model, consistent with the spatial bias analysis in Section 4.2.1 and further supporting the selection of the 30-day configuration as the primary model.

\subsection{Temporal Dynamics of Domain-Averaged LAI}
\label{subsec4.3}

While pixel-wise metrics provide a fundamental assessment of model accuracy, domain-averaged LAI time series provide a complementary perspective on model skill at the regional scale. By averaging across all valid pixels, domain-level LAI minimizes high-frequency spatial noise. Such regional mean behavior is important for sub-seasonal to seasonal forecasting and land surface modeling, where aggregated vegetation states serve as surface boundary conditions \citep{kumar2019}. We evaluate the continuous time series of domain-averaged LAI over the independent testing period (2023--2025).

\begin{figure}[H]
\centering
\includegraphics[width=1\textwidth]{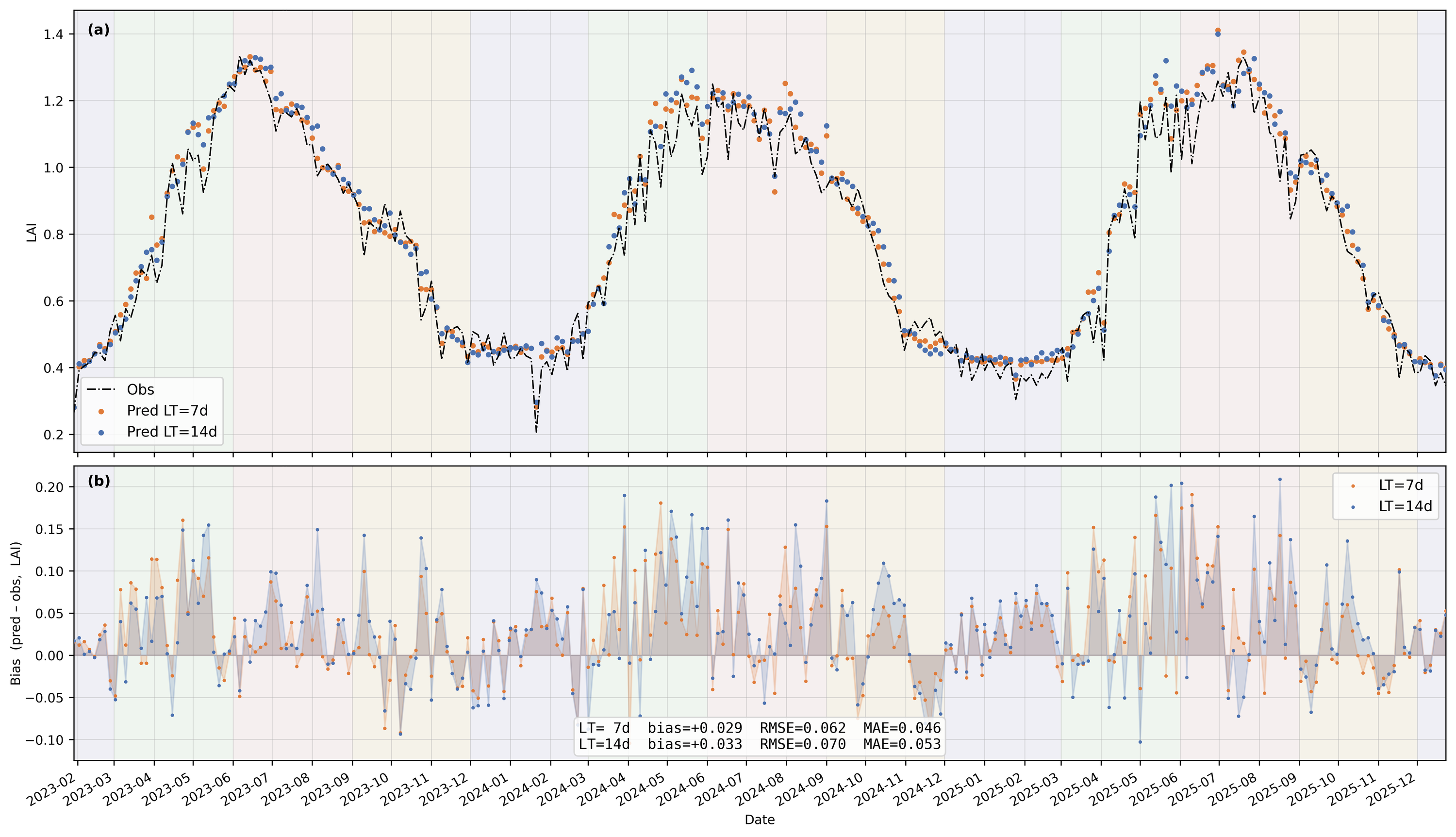}
\caption{Domain-averaged LAI time series from the 14-day model over the independent testing period (2023--2025). (a) Observed domain-averaged LAI (black dashed line) compared against model predictions at lead times of 7 days (orange dots) and 14 days (blue dots). Seasonal shading indicates DJF (blue), MAM (green), JJA (pink), and SON (yellow). (b) Temporal evolution of domain-averaged bias (predicted minus observed) at lead times of 7 days (orange) and 14 days (blue), with shaded areas indicating the bias magnitude. Summary statistics are reported in the lower right: mean bias, RMSE, and MAE of the domain-averaged LAI time series for each lead time, computed over the full evaluation period. Domain averages are computed over all quality-controlled valid pixels within the study domain.}
\label{fig:timeseries_domain_avg_14d}
\end{figure}

\begin{figure}[H]
\centering
\includegraphics[width=1\textwidth]{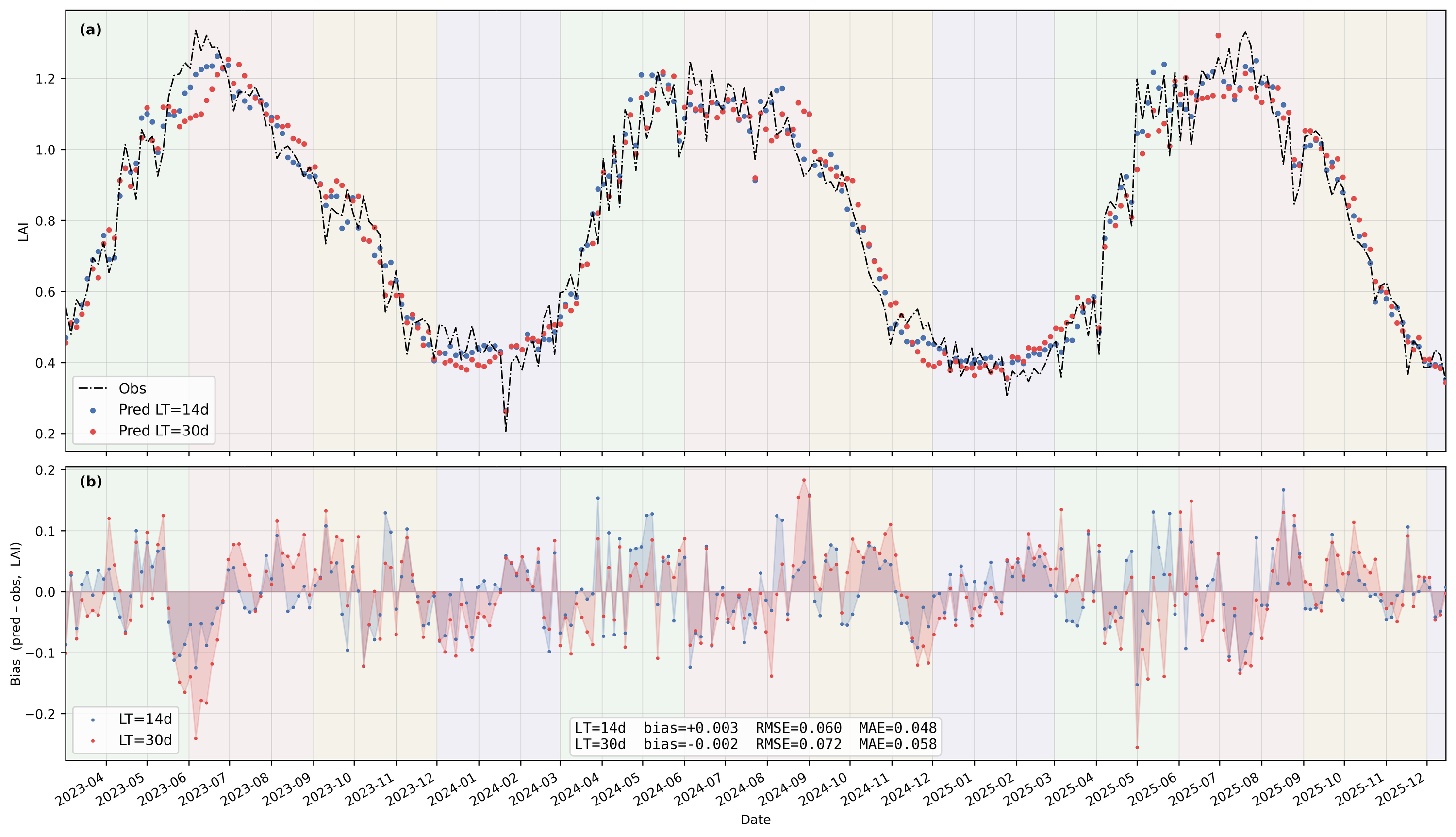}
\caption{Domain-averaged LAI time series from the 30-day model over the independent testing period (2023--2025). (a) Observed domain-averaged LAI (black dashed line) compared against model predictions at lead times of 14 days (blue dots) and 30 days (red dots). Seasonal shading indicates DJF (blue), MAM (green), JJA (pink), and SON (yellow). (b) Temporal evolution of domain-averaged bias (predicted minus observed) at lead times of 14 days (blue) and 30 days (red), with shaded areas indicating the bias magnitude. Summary statistics are reported in the lower right: mean bias, RMSE, and MAE of the domain-averaged LAI time series for each lead time, computed over the full evaluation period. Domain averages are computed over all quality-controlled valid pixels within the study domain.}
\label{fig:timeseries_domain_avg_30d}
\end{figure}

Figure~\ref{fig:timeseries_domain_avg_14d} and Figure~\ref{fig:timeseries_domain_avg_30d} show the domain-averaged LAI time series for the 14-day and 30-day models, respectively, compared against the MODIS-derived domain-averaged LAI as the observational reference. While the 30-day model shows lower spatial bias overall, the 14-day model captures the domain-averaged LAI more accurately during the active growing season of May--August 2023 and reproduces the elevated LAI variability observed in late July 2025, despite the two models showing nearly identical RMSE of the domain-averaged LAI time series (0.062 vs. 0.060 at a 14-day lead time) --- a negligible difference that masks meaningful differences in temporal dynamics. Spatial averaging over a large domain also suppresses local anomalies, potentially masking model behavior at finer spatial scales. These results suggest that the 30-day model is the preferred configuration as the primary model; however, for applications requiring greater sensitivity to temporal variability or deviations from the regional mean response, the 14-day model may offer advantages that aggregate metrics alone do not reveal. Further site-specific evaluation is recommended before applying either configuration for such purposes.

\subsection{Phenological Variability Across Vegetation Types}
\label{subsec4.4}

LAI magnitudes vary substantially across vegetation types. Forests generally sustain the highest LAI values, while shrublands and grasslands maintain comparatively lower values in the semi-arid and arid environments. Vegetation types also exhibit distinct seasonal dynamics driven by different climate controls and management regimes. Shrubs in the central-western portion of the domain follow a monsoon-driven phenology: LAI remains suppressed through the dry spring, reaches a pre-monsoon minimum in early July, then rises sharply with the onset of summer convective precipitation, sustaining peak greenness from JJA into September \citep{warter2023}. Grasslands across the domain initiate green-up earlier in spring, with the start of season controlled by a combination of temperature and soil moisture. In contrast to natural vegetation types, cropland LAI dynamics are largely governed by agricultural practices such as planting calendars and harvest timing, resulting in sharp increases and abrupt declines in LAI \citep{amin2021}.

\begin{figure}[H]
\centering
\includegraphics[width=1\textwidth]{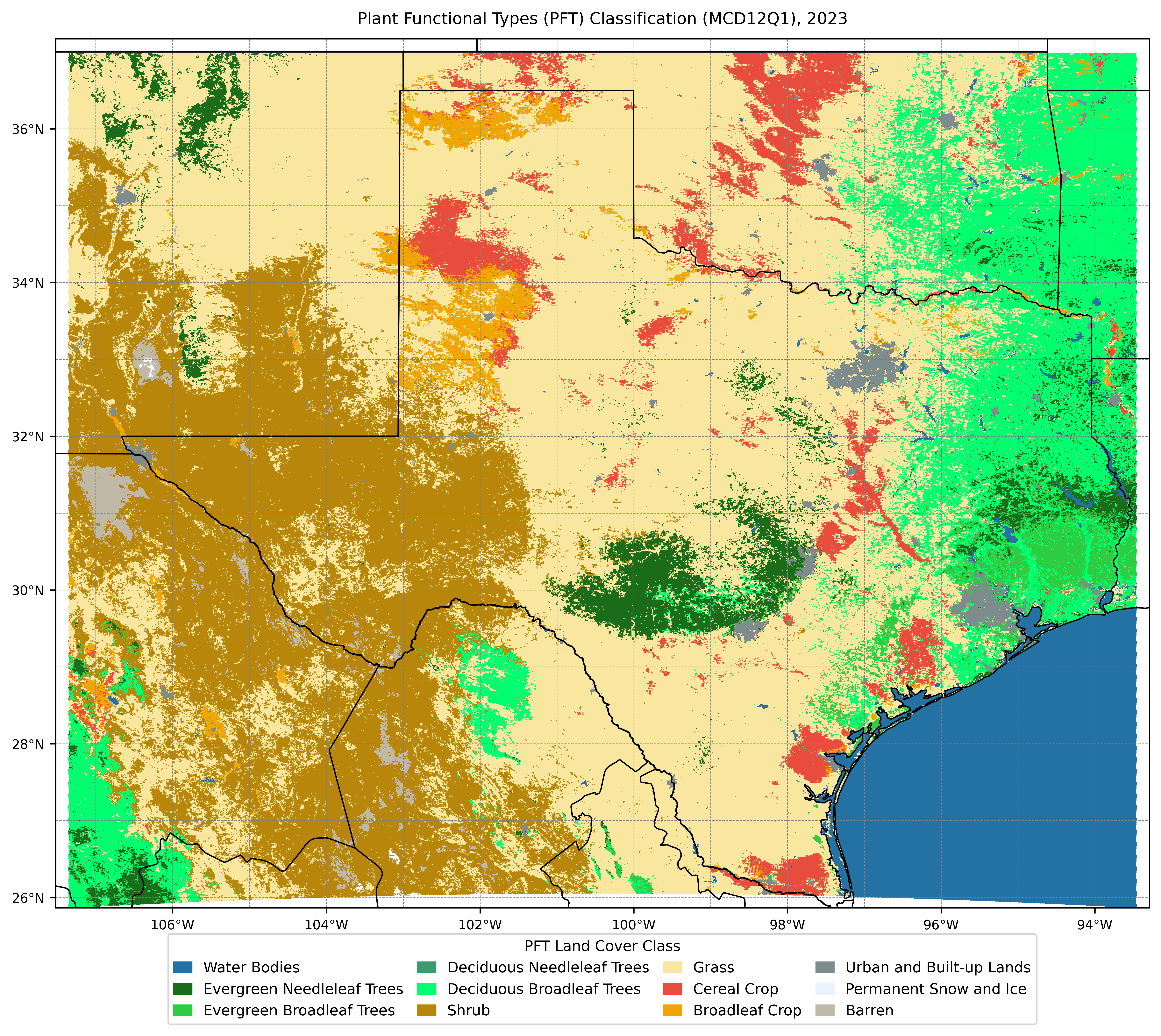}
\caption{Plant functional type (PFT) classification derived from the MODIS MCD12Q1 LC\_Type5 product for 2023.}
\label{fig:lc_pft_map}
\end{figure}

Figure~\ref{fig:lc_pft_map} shows the plant functional type (PFT) map derived from the MODIS MCD12Q1 product for 2023 \citep{Sulla-Menashe2022}, and Table~\ref{tab:pft_composition} summarizes the corresponding land cover composition of the study domain. Grasses (46.43\%), shrubs (20.81\%), and deciduous broadleaf trees (11.16\%) are the three dominant PFTs across the study domain, reflecting the semi-arid to sub-humid vegetation gradient from the Chihuahuan Desert through the Southern Plains to the eastern deciduous transition. Unlike other common land cover classification schemes, such as the International Geosphere-Biosphere Programme (IGBP) scheme and the University of Maryland (UMD) scheme, the PFT scheme distinguishes two crop classes --- cereal crops and broadleaf crops --- based on their distinct canopy structures and phenological behaviors. The following evaluation focuses on model performance across the major PFT classes in the domain.

\begin{table}[H]
\centering
\caption{Land cover composition of the study domain (MCD12Q1 PFT scheme) for 2023.}
\label{tab:pft_composition}
\begin{tabular}{clrr}
\hline
\textbf{Class} & \textbf{Label} & \textbf{Pixels} & \textbf{Percent} \\
\hline
0  & Water Bodies                  & 135,181   & 8.41\%  \\
1  & Evergreen Needleleaf Trees    &  61,840   & 3.85\%  \\
2  & Evergreen Broadleaf Trees     &  25,443   & 1.58\%  \\
3  & Deciduous Needleleaf Trees    &       2   & 0.00\%  \\
4  & Deciduous Broadleaf Trees     & 179,246   & 11.16\% \\
5  & Shrub                         & 334,320   & 20.81\% \\
6  & Grass                         & 745,942   & 46.43\% \\
7  & Cereal Crop                   &  61,425   & 3.82\%  \\
8  & Broadleaf Crop                &  31,040   & 1.93\%  \\
9  & Urban and Built-up Lands      &  17,838   & 1.11\%  \\
10 & Permanent Snow and Ice        &      14   & 0.00\%  \\
11 & Barren                        &  14,434   & 0.90\%  \\
\hline
   & \textbf{Total valid pixels}   & \textbf{1,606,725} & \textbf{100.00\%} \\
\hline
\end{tabular}
\end{table}

\subsubsection{Model Performance over Forests}
\label{subsec4.4.1}

\begin{figure}[H]
\centering
\includegraphics[width=1\textwidth]{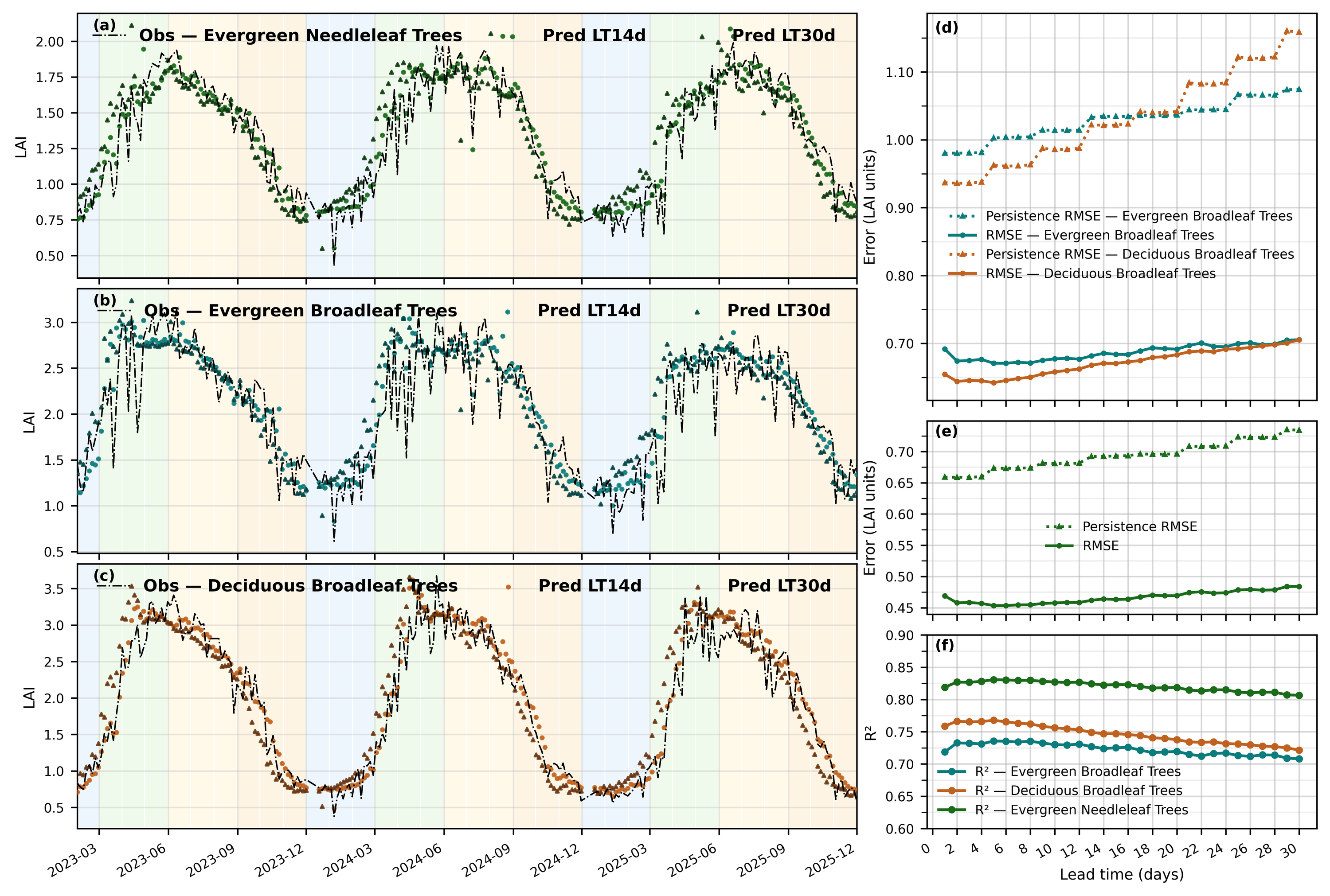}
\caption{Model evaluation for tree plant functional types. Left panels show domain-averaged LAI time series (2023--2025) for (a) evergreen needleleaf trees, (b) evergreen broadleaf trees, and (c) deciduous broadleaf trees, with observations (dashed line), 14-day lead time predictions (LT14d, circles), and 30-day lead time predictions (LT30d, triangles). Seasonal shading indicates DJF (blue), MAM (green), JJA (red), and SON (orange). Right panels show skill metrics as a function of lead time (1--30 days) for all three tree types: (d) RMSE (solid lines) and persistence baseline RMSE (dotted lines) in LAI units, and (e) coefficient of determination ($R^{2}$).}
\label{fig:landtype_domainavg_tree}
\end{figure}

Figure~\ref{fig:landtype_domainavg_tree} presents model performance across three primary tree classifications: deciduous broadleaf, evergreen needleleaf, and evergreen broadleaf, which account for approximately 12\%, 4\%, and 2\% of the study domain in 2023 and 2024. The deciduous needleleaf class is excluded, as it comprises less than 0.0002\% of the domain. The model outperforms the persistence forecast for all three tree types, with $R^2$ exceeding 0.7 across all lead times. The model successfully captures both seasonal and interannual variability across all tree types, accurately predicting the anomalously low LAI values observed in January 2024 even at a 30-day lead time.

\subsubsection{Model Performance Over Grasslands}
\label{subsec4.4.2}

\begin{figure}[H]
\centering
\includegraphics[width=1\textwidth]{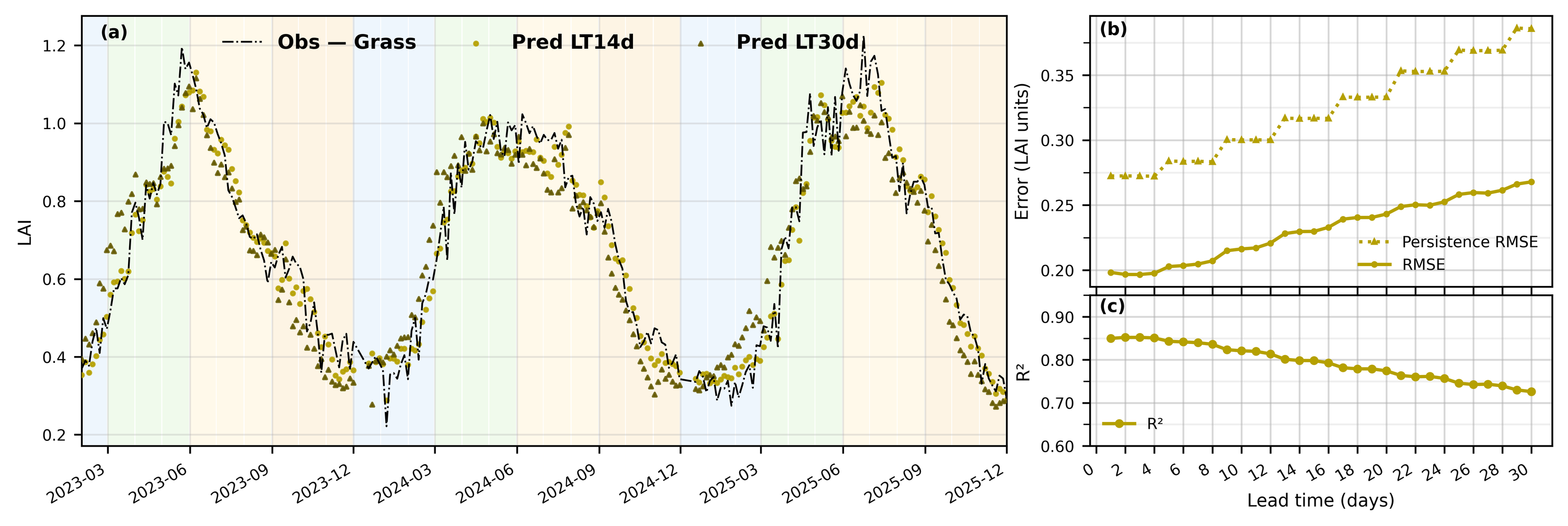}
\caption{Model evaluation for grasslands. (a) Domain-averaged LAI time series from 2023--2025, with observations (dashed line), 14-day lead time predictions (LT14d, circles), and 30-day lead time predictions (LT30d, triangles). Seasonal shading: DJF (blue), MAM (green), JJA (red), and SON (orange). (b) RMSE (solid line) and persistence baseline RMSE (dotted line) in LAI units as a function of lead time (1--30 days). (c) Coefficient of determination ($R^{2}$) as a function of lead time.}
\label{fig:landtype_grass}
\end{figure}

Grasslands constitute the dominant land cover type across the study domain, and the model demonstrates robust predictive skill over these regions. As illustrated in Figure~\ref{fig:landtype_grass}, the model maintains a high $R^2$ of 0.73 even at a lead time of 30 days. The time series shows several notable phenological anomalies: a sharp increase in LAI in May 2023, a sudden LAI decline in January 2024, and sustained peak greenness throughout June and July of 2025. Both the 14-day and 30-day forecasts successfully capture the abrupt LAI decline in January 2024, and gradually adjust to the rapid spring green-up in 2023. However, while the 14-day forecast effectively tracks the sustained high LAI in early summer 2025, the 30-day forecast struggles to capture the magnitude of this trend. 

\begin{figure}[H]
\centering
\includegraphics[width=1\textwidth]{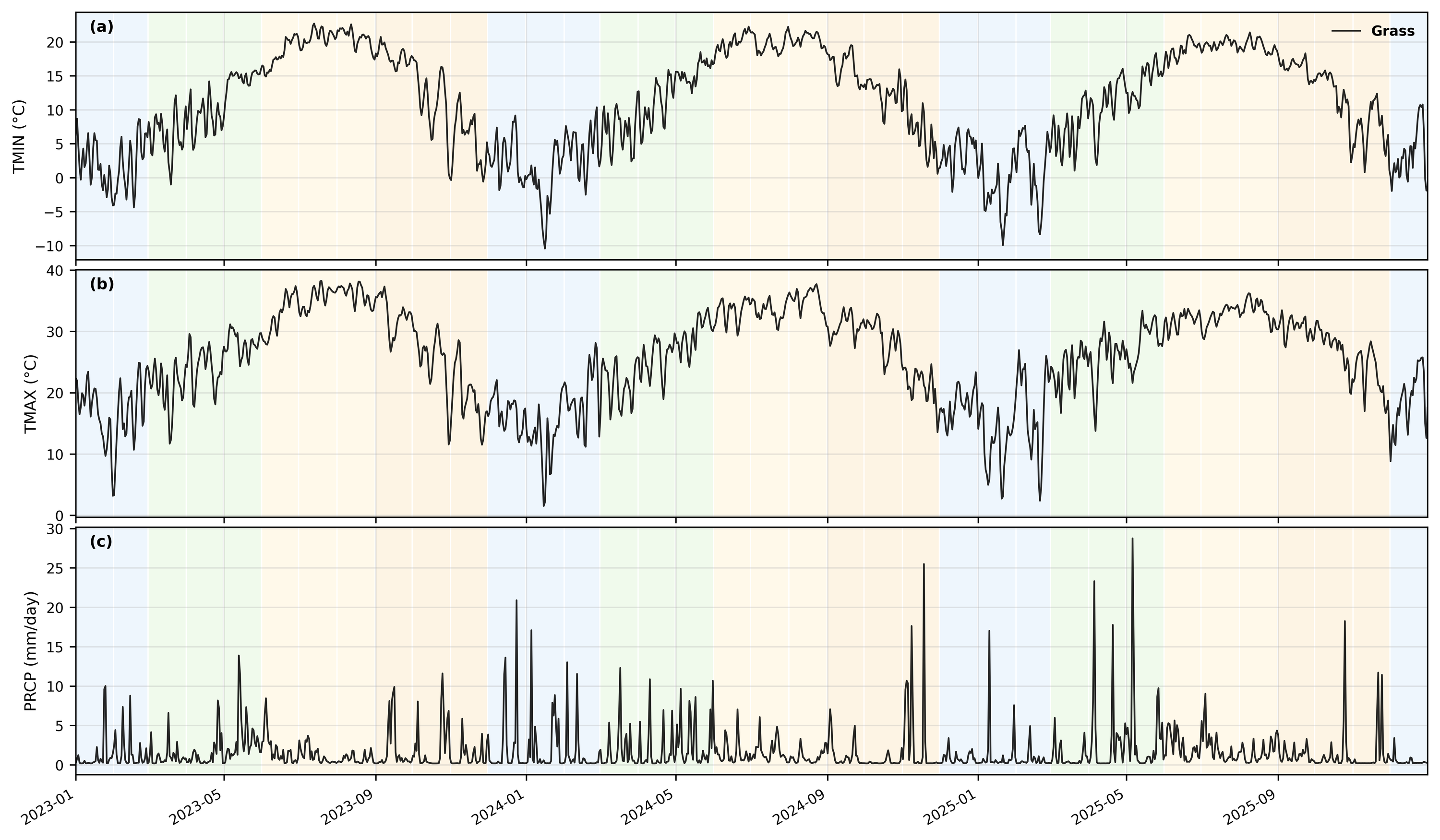}
\caption{Domain-averaged meteorological forcing time series over grassland pixels (2023--2025): (a) daily minimum temperature (T$_{\text{min}}$, $^{\circ}$C), (b) daily maximum temperature (T$_{\text{max}}$, $^{\circ}$C), and (c) daily precipitation (PRCP, mm day$^{-1}$). Seasonal shading indicates DJF (blue), MAM (green), JJA (yellow), and SON (orange).}
\label{fig:weather_grass_timeseries}
\end{figure}

Because grasses are shallow-rooted and highly sensitive to recent precipitation pulses \citep{ogle2004}, examining the underlying weather forcing provides important context for this behavior. Figure~\ref{fig:weather_grass_timeseries} presents the domain-averaged daily maximum temperature (Tmax), minimum temperature (Tmin), and precipitation across the grassland regions. In mid-May 2023, the region experienced a notably higher precipitation pulse compared to the same period in 2024 and 2025, while temperatures were comparable across years. This moisture influx was likely one of the drivers of the sharp LAI increase observed in May 2023. Because 30-day forecasts initialized prior to this event lacked information about the subsequent precipitation pulse, green-up was initially underestimated; forecasts initialized closer to the event correctly captured the LAI response. Similarly, the sustained high LAI observed in June and July 2025 correlates strongly with higher precipitation and cooler Tmax values relative to 2023 and 2024. The 30-day forecast for June, driven primarily by the drier meteorological conditions present at initialization in May, predictably underestimated this growth. In contrast, 14-day forecasts initialized in July, when the wetter conditions were already established, successfully captured the elevated LAI.

\subsubsection{Model Performance Over Shrubs}
\label{subsec4.4.3}

\begin{figure}[H]
\centering
\includegraphics[width=1\textwidth]{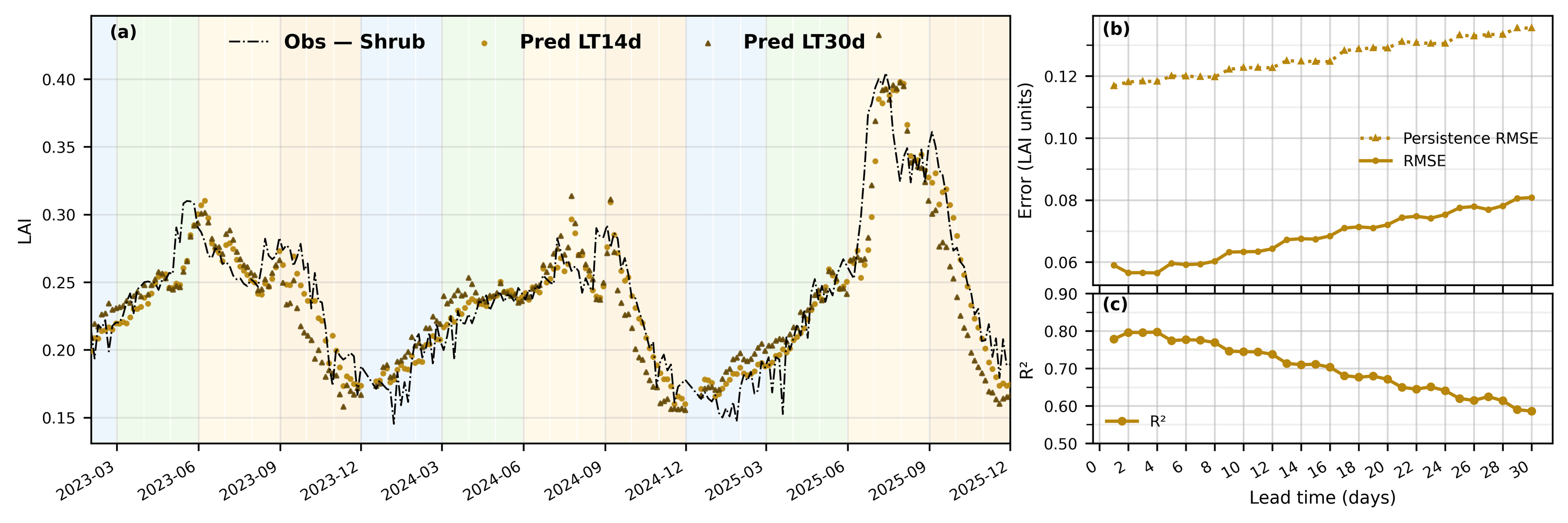}
\caption{Model evaluation for shrublands. (a) Domain-averaged LAI time series from 2023--2025, with observations (dashed line), 14-day lead time predictions (LT14d, circles), and 30-day lead time predictions (LT30d, triangles). Seasonal shading: DJF (blue), MAM (green), JJA (yellow), and SON (orange). (b) RMSE (solid line) and persistence baseline RMSE (dotted line) in LAI units as a function of lead time (1--30 days). (c) Coefficient of determination ($R^{2}$) as a function of lead time.}
\label{fig:landtype_shrub}
\end{figure}

Figure~\ref{fig:landtype_shrub} presents the model's predictive skill over shrublands, which primarily occupy the western portion of the domain. These ecosystems exhibit high seasonality, with a short period of intense greening shortly after the onset of North American Monsoon (NAM) precipitation \citep{forzieri2011}. Observational LAI time series highlight a contrast with trees and grasses: shrubland green-up is distinctly delayed, remaining suppressed through the spring and peaking from summer into early autumn. The 14-day forecast reproduces this dynamic, achieving an $R^2$ of approximately 0.70 and successfully capturing both the delayed onset and the rapid green-up across all three testing years.

\begin{figure}[H]
\centering
\includegraphics[width=1\textwidth]{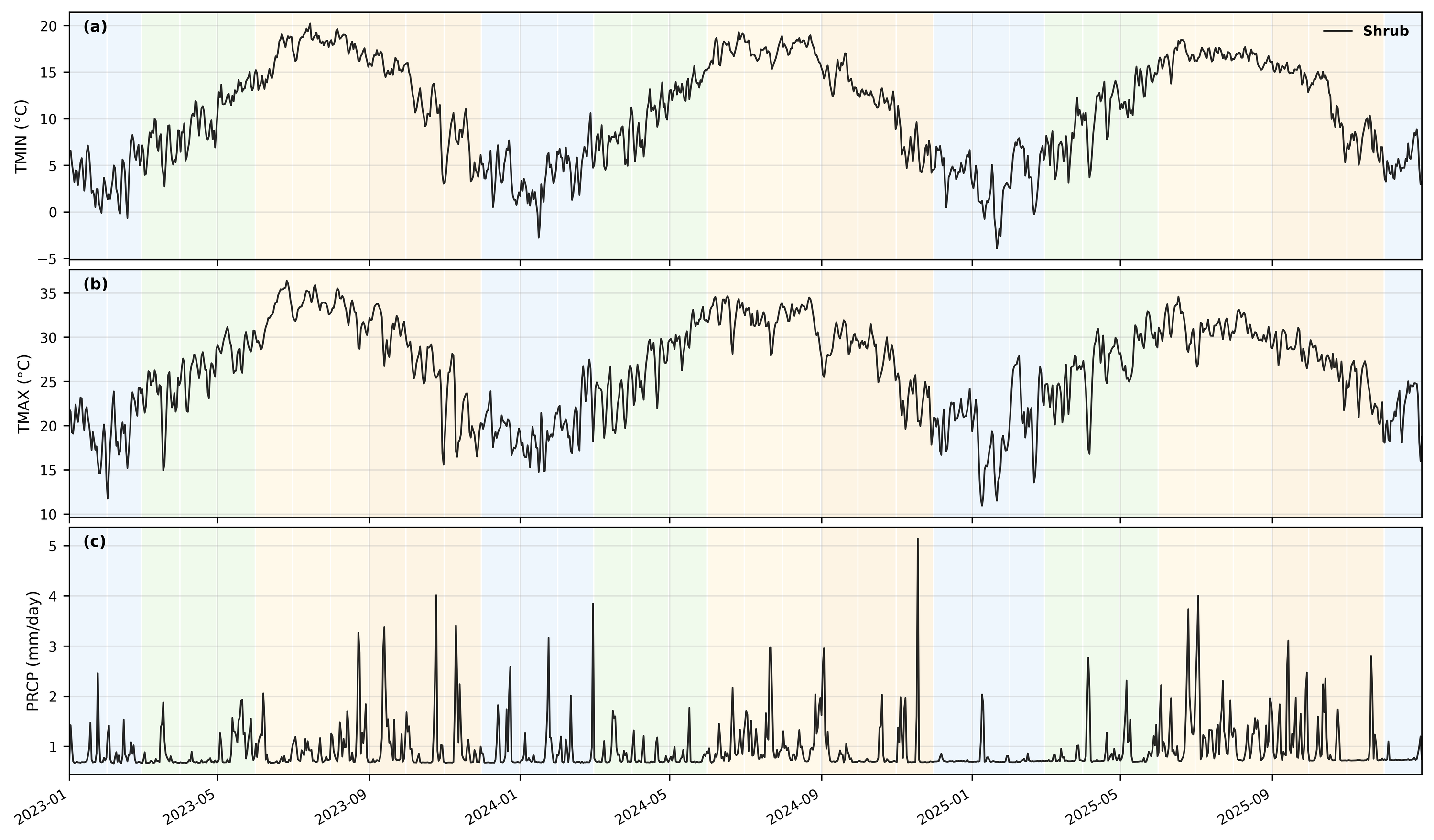}
\caption{Domain-averaged meteorological forcing time series over shrub pixels (2023--2025): (a) daily minimum temperature (T$_{\text{min}}$, $^{\circ}$C), (b) daily maximum temperature (T$_{\text{max}}$, $^{\circ}$C), and (c) daily precipitation (PRCP, mm day$^{-1}$). Seasonal shading indicates DJF (blue), MAM (green), JJA (yellow), and SON (orange).}
\label{fig:weather_shrub_timeseries}
\end{figure}

Performance degrades at the 30-day lead time ($R^2 \approx 0.65$), primarily due to a tendency to overestimate early-season LAI. Notably, in February and March 2024, the 30-day forecast predicted LAI values 0.025--0.05 higher than observations, before realigning with the observed trajectory in April. Figure~\ref{fig:weather_shrub_timeseries} shows that the winter of 2023--2024 was wet; because the 30-day forecasts for early spring were initialized using these wet conditions, the model prematurely triggered the green-up phase. The 30-day forecasts for April were initialized using March conditions, which reflected average precipitation, and the predicted trajectory moved closer to the observed delayed green-up, though not fully converging with observations.

Performance degrades at the 30-day lead time ($R^2 \approx 0.65$), primarily due to a tendency to overestimate early-season LAI. Notably, in February and March 2024, the 30-day forecast predicted LAI values 0.025--0.05 higher than observations, before realigning with the observed trajectory in April. Figure~\ref{fig:weather_shrub_timeseries} indicates that the winter of 2023--2024 was wet; because the 30-day forecasts for early spring were initialized using these wet conditions, the model prematurely triggered the green-up phase. The 30-day forecasts for April were initialized using March conditions, which reflected average precipitation, and the predicted trajectory moved closer to the observed delayed green-up, though not fully converging with observations.

Overall, shrublands exhibit the lowest 30-day $R^2$ among the natural vegetation classes. A critical factor worth further investigation is the quality and spatial density of the precipitation input data. The western domain features a notably sparser network of precipitation stations (Figure~\ref{fig2}). Given the high sensitivity of desert shrubs to localized precipitation pulses, sparse station coverage likely introduces uncertainty into the model initializations, constraining predictability at longer lead times.

\subsubsection{ Model Performance Over Croplands}
\label{subsec4.4.4}

In contrast to natural vegetation, cropland phenology is primarily governed by human activities rather than climate forcing alone. A major advantage of this machine learning framework is its capacity to implicitly learn and generalize these human-driven patterns --- such as abrupt harvest events --- directly from historical LAI changes. This remains a significant challenge for traditional, physics-based vegetation dynamics models.

\begin{figure}[H]
\centering
\includegraphics[width=1\textwidth]{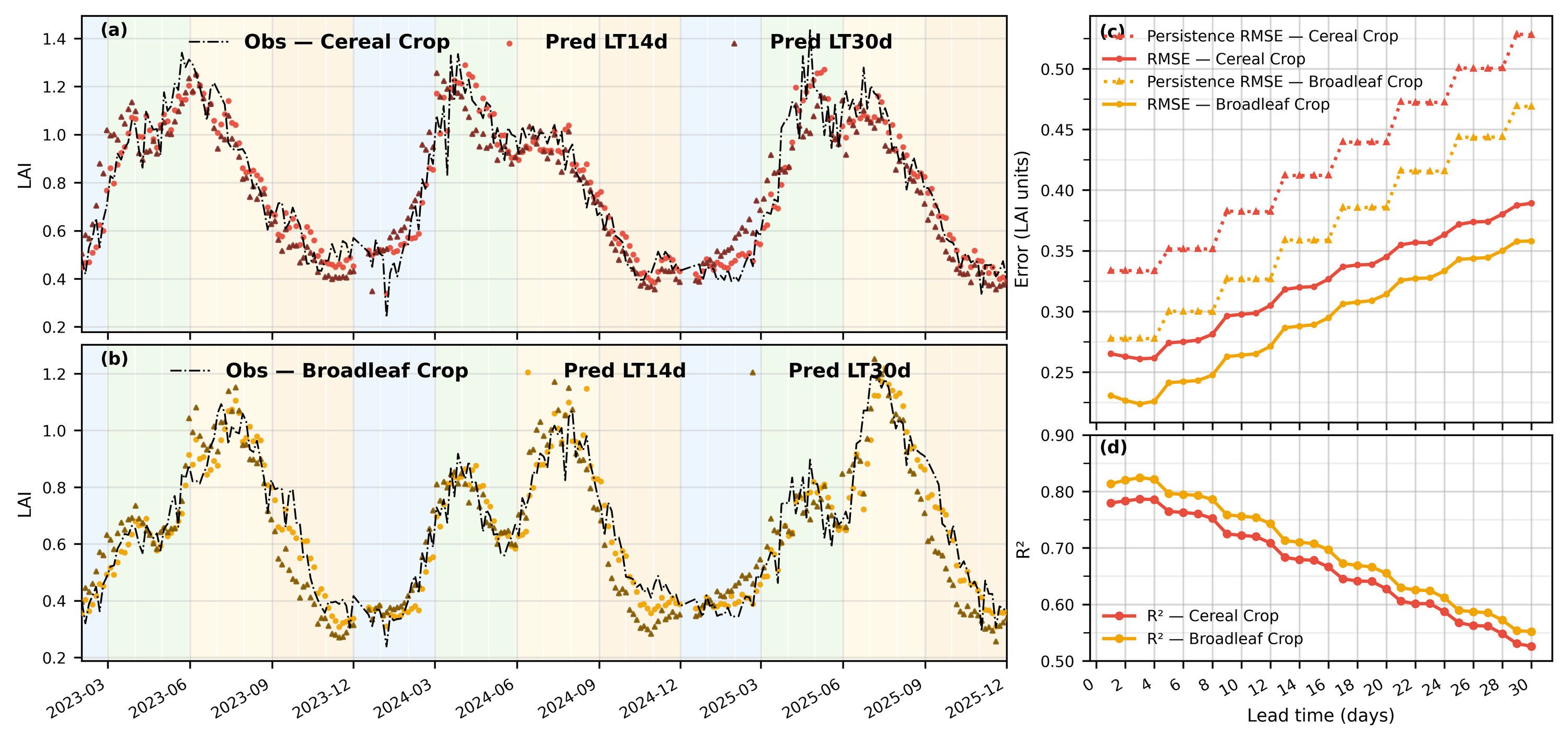}
\caption{Model evaluation for cropland plant functional types. Left panels show domain-averaged LAI time series from 2023 to 2025 for (a) cereal crops and (b) broadleaf crops, with observations (dashed line), 14-day lead time predictions (LT14d, circles), and 30-day lead time predictions (LT30d, triangles). Seasonal shading indicates DJF (blue), MAM (green), JJA (yellow), and SON (orange). Right panels show skill metrics as a function of lead time (1--30 days) for both crop types: (c) RMSE (solid lines) and persistence baseline RMSE (dotted lines) in LAI units, and (d) coefficient of determination ($R^{2}$).}
\label{fig:landtype_crop}
\end{figure}

Figure~\ref{fig:landtype_crop} presents the LAI forecasts and associated metrics for both cereal and broadleaf crops. Both crop types show meaningful forecast skill at a 30-day lead time, with $R^2$ ranging from 0.53 to 0.55, though performance is more variable than for natural vegetation types, reflecting the added complexity of human-driven management practices. Both 14-day and 30-day forecasts reproduce major interannual phenological shifts and accurately capture the dual LAI reductions associated with spring and summer harvest windows for both crop types. However, the forecasts struggle to resolve minor fluctuations, such as the subtle LAI variations observed in cereal crops from late autumn to winter. These deviations are likely due to the framework's spatial limitations when resolving small-scale agricultural parcels, a factor that warrants further investigation.

\section{Discussion}
\label{sec5}

The sequence-to-sequence ConvLSTM architecture demonstrates predictive skill up to a 30-day horizon at 1-km resolution, achieving a domain-averaged RMSE of approximately 0.36, which outperforms the persistence baseline RMSE of 0.56. Beyond aggregate error metrics, the model maintains spatial consistency across the domain and reproduces key phenological features across vegetation types and seasons, supporting its potential utility as an LAI forcing in land surface and climate models.

\begin{figure}[H]
\centering
\includegraphics[width=1\textwidth,height=0.85\textheight,keepaspectratio]{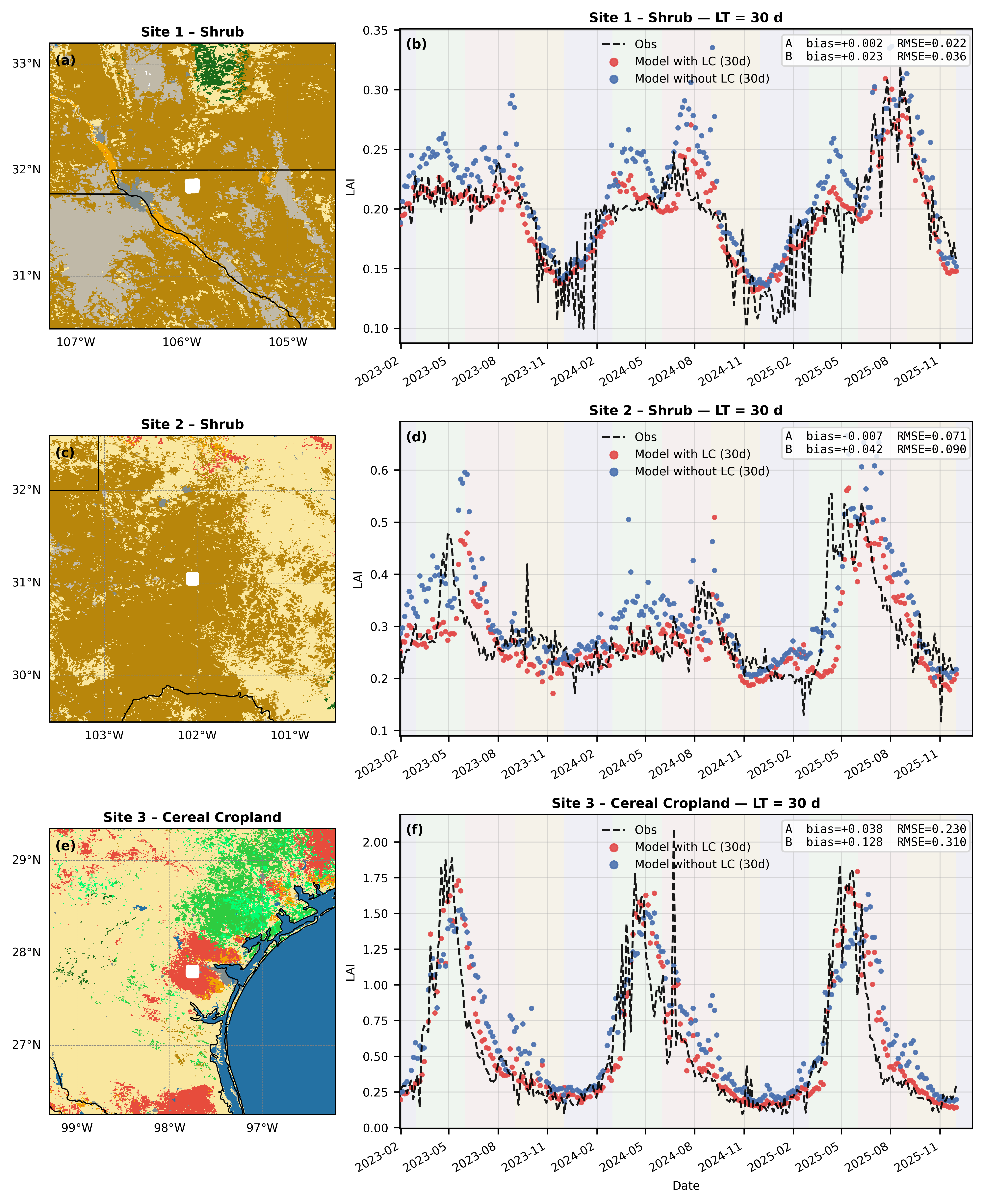}
\caption{Site-level LAI comparison between the model with land cover predictors (Model A, red) and without land cover predictors (Model B, blue) at a 30-day lead time. Left panels (a, c, e) show the spatial extent and land cover composition of each representative site (white rectangles); the PFT color legend is shown in Figure~\ref{fig:lc_pft_map}. Right panels (b, d, f) show the domain-averaged observed LAI (black dashed) alongside Model A and Model B predictions. Bias and RMSE values for each model configuration are annotated in each panel.}
\label{fig:landcover_ablation}
\end{figure}

LAI dynamics are strongly land-cover dependent. As demonstrated in Section 4.4, different vegetation types exhibit distinct phenological responses to meteorological forcing --- shrublands show delayed green-up dynamics, and croplands are strongly modulated by agricultural management. Incorporating dynamic land cover as a time-varying input allows the model to learn PFT-specific LAI response patterns rather than averaging across heterogeneous vegetation, which is particularly important when targeting broad domains such as CONUS. Figure~\ref{fig:landcover_ablation} presents a site-level LAI analysis comparing the model with and without land cover predictors. Representative local regions of approximately 0.1$^{\circ}$ $\times$ 0.1$^{\circ}$ ($\sim$121 km$^{2}$) are examined. Two shrubland sites in the Chihuahuan Desert ecoregion and one cropland site along the southeast Texas coast are selected. For the shrubland sites, the model with land cover predictors better captures the delayed late-summer green-up, with predicted LAI values more closely following observed ranges and without anomalously large values. For the cropland site, the model with land cover predictors captures the LAI decline associated with corn and sorghum harvest beginning in July, while the model without land cover predictors overestimates LAI during this period. Domain-averaged and PFT-stratified metrics at a 30-day lead time, however, remain relatively close between the two configurations, as summarized below:

\begin{table}[H]
\centering
\caption{Domain-averaged and PFT-stratified performance metrics at a 30-day lead time, with and without land cover (LC) as a model predictor.}
\label{tab:landcover_ablation}
\begin{tabular}{lcccccc}
\toprule
Configuration & RMSE & MAE & $R^2$ & RMSE$_{\text{shrub}}$ & RMSE$_{\text{cereal}}$ & RMSE$_{\text{broadleaf}}$ \\
\midrule
Model with LC    & 0.361 & 0.194 & 0.823 & 0.081 & 0.389 & 0.358 \\
Model without LC & 0.370 & 0.205 & 0.814 & 0.092 & 0.412 & 0.376 \\
\bottomrule
\end{tabular}
\end{table}

This reflects a common trade-off in machine learning model design --- that architecture complexity and additional predictors do not always yield proportional improvements in aggregate performance. The improvement from incorporating land cover predictors is not uniform: it benefits vegetation types with distinctive phenology, such as shrublands and croplands, while the domain-averaged gain remains modest. From a practical standpoint, the model without land cover predictors offers a simpler input pipeline, which can be an advantage in near-real-time forecasting contexts. The choice between configurations therefore depends on the target application --- where PFT-specific accuracy is required, the additional complexity is justified; where domain-wide LAI forecasting skill is the primary goal, the simpler configuration remains competitive. Future sensitivity experiments within land surface and climate modeling frameworks will be needed to quantify whether these finer-scale RMSE improvements translate into better forecasting skill once the LAI fields are ingested into such models.

Several limitations warrant consideration. Forecast skill over shrublands is limited compared to other vegetation types, which may partly reflect the effective spatial resolution of the interpolated meteorological variables, particularly given the sparser precipitation station coverage across the western domain. Additionally, as with other autoregressive sequence-to-sequence architectures, prediction errors can accumulate across the forecast horizon, a limitation that becomes more pronounced at longer lead times. Building on this proof-of-concept over the South-Central US, future work will extend the framework toward CONUS-wide application, supporting the broader goal of operationally relevant LAI forecasting.

\section*{Acknowledgments}
This research was supported by the NOAA Weather Program Office (WPO) Subseasonal to Seasonal (S2S) Program under Grant NA23OAR4590386, the NASA S2S Program under Award 80NSSC23K0504, and the CSU Scott Foundation High-Impact Research Fund.

\let\oldbibitem\bibitem
\renewcommand{\bibitem}{\setlength{\itemsep}{0.5\itemsep}\oldbibitem}
\bibliographystyle{elsarticle-harv}
\bibliography{references}

\end{document}